\documentclass[12pt,preprint]{aastex}
\usepackage{graphicx}
\usepackage{natbib}

\def\Msolar{M$_\odot$}
\def\msun{M$_\odot$}
\def\mstar{M$_\star$}
\def\rinit{r$_{\rm init}$}
\def\vinit{v$_{\rm init}$}
\def\vej{v$_{\rm ej}$}
\def\vrad{v$_{\rm rad}$}
\def\deg{$^{\rm o}$}
\def\kms{km~s$^{-1}$}
\def\GC{Galactic Center}

\begin{document}

\title{Runaway Stars, Hypervelocity Stars, and Radial Velocity Surveys}

\author{Benjamin C. Bromley}
\affil{Department of Physics, University of Utah,
 115 S 1400 E, Rm 201, Salt Lake City, UT 84112}
\email {bromley@physics.utah.edu}

\author{Scott J. Kenyon}
\affil{Smithsonian Astrophysical Observatory,
 60 Garden St., Cambridge, MA 02138}
\email{skenyon@cfa.harvard.edu}

\author{Warren R. Brown}
\affil{Smithsonian Astrophysical Observatory,
 60 Garden St., Cambridge, MA 02138}
\email{wbrown@cfa.harvard.edu}

\author{Margaret J. Geller}
\affil{Smithsonian Astrophysical Observatory,
 60 Garden St., Cambridge, MA 02138}
\email{mgeller@cfa.harvard.edu}

\clearpage

\begin{abstract}

	Runaway stars ejected from the Galactic disk populate the halo of the 
Milky Way.  To predict the spatial and kinematic properties of runaways, we 
inject stars into a Galactic potential, compute their trajectories through the 
Galaxy, and derive simulated catalogs for comparison with observations.  Runaways 
have a flattened spatial distribution, with higher velocity stars at Galactic 
latitudes less than 30\deg. Due to their shorter stellar lifetimes, massive 
runaway stars are more concentrated towards the disk than low mass runaways.  
Bound (unbound) runaways that reach the halo probably originate from distances 
of 6--12 kpc (10--15 kpc) from the Galactic center, close to the estimated origin 
of the unbound runaway star HD 271791.  Because runaways are brighter and have 
smaller velocities than hypervelocity stars (HVSs), radial velocity surveys are 
unlikely to confuse runaway stars with HVSs. We estimate that at most 1 runaway 
star contaminates the current sample. We place an upper limit of 2\% on the 
fraction of A-type main sequence stars ejected as runaways.

\end{abstract}


\section{Introduction}

	Radial velocity surveys of the Milky Way halo can now reveal rare 
velocity outliers resulting from dynamical processes throughout the Galaxy.  
Presently there are three known categories of velocity outliers among main 
sequence stars: (1) runaway stars originating in the Galactic disk 
\citep{humason47}, (2) velocity outliers from tidal disruption of dwarf 
satellites within the halo \citep{ibata94}, and (3) hypervelocity stars 
(HVSs) originating in the Galactic center \citep{brown05}.  Recent
simulations predict the observable spatial and velocity distributions for
HVSs \citep{kenyon08} and for remnants of tidal disruption \citep{abadi08}.  
Although runaway stars have been known for a long time, there has been only 
limited investigation of their observable spatial/velocity distribution
throughout the Galaxy.  

Here, we simulate the production of runaways and trace their orbits through 
the Galaxy, concentrating on runaways that reach the halo.  We derive the 
spatial and velocity distributions of runaways as a function of mass and 
use these results to predict observable quantities for comparison with data 
from radial velocity surveys. From these results, we estimate the fraction
of runaways in surveys of halo stars. We also evaluate the likelihood of
runaway star contaminants in targeted surveys for HVSs.

\subsection {History and Challenges}

First reported by \citet{humason47}, runaway stars are short-lived stars at 
unexpectedly large distances and velocities relative to their probable site 
of origin  \citep{blaauw54, greenstein57}.  In the solar neighborhood, roughly 
10--30\% of O stars and 5--10\% of B stars are runaways \citep{gies87,stone91}. 
The main population of runaways has a velocity dispersion, 30 \kms, roughly 3 times 
larger than the velocity dispersion of non-runaway stars in the Galactic disk. 
However, several runaways have velocities $\gtrsim$ 100 \kms\ relative to the local 
standard of rest \citep[e.g.,][]{gies87,stone91,martin06} and large distances, 
$\gtrsim$ 500 pc, from the Galactic plane \citep{green74,martin06}.  The large 
velocity dispersion and vertical scale height suggests a dynamical process that 
ejects runaways from the thin disk into the halo.

Runaways probably originate within star-forming regions in the disk of the
Milky Way.  Disruption of massive binaries is the likely source of runaways. 
In the binary supernova mechanism, a main sequence runaway is ejected when 
its former companion star explodes as a supernova; subsequent mass loss and
the kick velocity from the asymmetric explosion are sufficient to unbind
the binary \citep[see][]{blaauw61,hills83}.  In this scenario, the runaway 
moves away with roughly the sum of the kick velocity and the orbital velocity 
of the binary, a velocity physically limited by the orbital velocity at the 
surface of the stars.  In the dynamical ejection mechanism, runaways are 
ejected in dynamical three- or four-body interactions; the outcome is any 
combination of single stars and binaries \citep{poveda67,hoffer83,hut83}.  The 
maximum ejection velocity from dynamical binary-binary encounters is formally 
the escape velocity of the most massive star \citep{leonard91}.  

Understanding whether the proposed formation mechanisms can produce the observed 
runaway star populations requires high quality kinematic data and a clear 
understanding of the dynamics of close binaries and dense star clusters. For large 
ensembles of runaways, the low binary frequency of runaways favors the dynamical 
ejection mechanism \citep{gies86}. However, the predicted kick velocity of supernovae 
is uncertain \citep[e.g.,][]{burrows95,murphy04}; if it is large enough, runaways 
produced by the supernova ejection mechanism will also be single.  For individual runaways, 
it is possible to identify a point of origin by measuring accurate distances, proper 
motions, and radial velocities \citep[e.g.,][]{hoogerwerf01,dewit05,martin06,heber08}. 
For example, using Hipparcos data and 
accurate radial velocities, \citet{hoogerwerf01} (i) associate the runaway star 
$\zeta$ Oph and the pulsar PSR J1932+1059 with a supernova in the Sco OB2 association 
and (ii) confirm that AE Aur, $\mu$ Col, and $\iota$ Ori were ejected from the 
Trapezium cluster in the Orion nebula. Thus, both runaway mechanisms occur in nature.

Existing observational and theoretical results place few constraints on the spatial and velocity 
distribution of runaways throughout the Milky Way.  Because accurate proper motions are often 
crucial for identifying runaways, known runaways are observed largely in the solar neighborhood 
\citep[e.g.][]{conlon90, holmgren92, mitchell98, rolleston99, hoogerwerf01, ramspeck01, magee01, 
lynn04, martin04, martin06}.  Although some runaways have large distances from the Galactic plane 
\citep[e.g.,][] {green74,heber08}, few are in the Galactic halo. Numerical simulations currently
provide little insight into runaways as probes of the Galactic halo structure.  \citet{davies02} 
simulated the distribution of runaways, but only in the context of high velocity white dwarfs near 
the Sun.  \citet{martin06} proposed using runaways to constrain the Galactic potential, but he 
concluded that known runaways provide few constraints on the large-scale properties of the Galaxy.

\subsection {From the Galactic Disk to the Halo}

Recent observations of `hypervelocity stars' (HVSs) in the halo motivate a
broader investigation of the spatial and velocity distribution of runaways.  
The first HVS is a 3 \Msolar\ main sequence star traveling at least twice the 
escape velocity of the Galaxy at its distance of $\sim$ 110 kpc \citep{brown05,brown09a}.  
After two other serendipitous discoveries \citep{hirsch05, edelmann05}, subsequent 
targeted searches of the halo yielded a sample of 15 unbound HVSs with Galactic
rest-frame velocities of 350--700 \kms\ and a similar number of bound HVSs with
rest-frame velocities of 275--350 \kms\ \citep{brown06,brown06b,brown07a,brown07b,brown09a}.  
Although the known HVSs are 3--4 \msun\ main sequence stars roughly uniformly 
distributed in Galactic latitude, they are not isotropically distributed on the 
sky \citep{brown09b}.

The large space velocity of HVS1 suggests an origin in the \GC. As first predicted
by \citet{hills88}, the tidal field of the massive black hole at the \GC\ can 
unbind a close binary and eject one of the stars at velocities exceeding 2000 \kms. 
To distinguish these high velocity stars from traditional runaway stars, Hills coined 
the term `hypervelocity star'.  For typical ejection velocities expected from this 
mechanism \citep{hills88}, we expect a range of velocities similar to those observed 
in HVSs.  Although other dynamical mechanisms involving a black hole can produce HVSs 
\citep[e.g.,][]{han03,yu03,ole08}, the Hills mechanism makes clear predictions for 
the expected velocity distribution of HVSs ejected from the \GC.  \citet{kenyon08} use 
these predictions in a numerical simulation of the trajectories of HVSs through the 
Galaxy. They show that the relative number of bound and unbound HVSs predicted by the 
Hills mechanism agrees with observations of known HVSs.

Recent observations suggest that runaways can also achieve unbound velocities.
The likely unbound star HD~271791 was probably ejected from the disk at 12--16 kpc
from the \GC\ \citep{heber08, przybilla08a}.  \citet{justham08} propose that 
the hot subdwarf US 708, with a heliocentric radial velocity of 708 \kms\
\citep{hirsch05}, is also a runaway. The apparent overlap in the velocity and
spatial distributions of runaways and HVSs suggests that several mechanisms may
inject massive main sequence stars into the halo. The relative contributions of
these processes to structure in the halo remains unknown.

	The distribution of runaways and HVSs is also an important issue for large
radial velocity surveys.  Surveys like RAVE \citep{zwitter08} and SEGUE
\citep{adelman08} measure radial velocities for hundreds of thousands of 
stars in the thin disk, thick disk, and halo.  These data will provide 
fundamental constraints on the escape velocity and total mass of the Milky Way 
\citep{smith07, xue08, siebert08}.  Although there are theoretical predictions for
observable properties of HVSs in the halo \citep[e.g.,][]{kenyon08}, there are no
predictions for runaway stars in the halo. Thus, the contribution of runaways to 
velocity outliers in these surveys is unknown.

	Here, we use numerical simulations to make a first assay of the spatial and 
kinematic signature of runaways in large radial velocity surveys.  Our focus is on 
intermediate mass 1.5--6~\Msolar\ main sequence stars ejected from the disk into the 
halo.  Intermediate mass stellar lifetimes are $10^8$ -- $10^9$ yr; thus, these stars 
formed recently in the disk.  The supernova binary disruption scenario requires that 
the former companion of the runaway was a more massive star with a shorter lifetime. 
We do not consider massive ($>6$ \Msolar) runaways, because they do not live long 
enough to reach large distances in the halo.  We also do not consider low mass stars 
because they are intrinsically faint and unobservable at large distances
\citep{kollmeier07,kenyon08}.  

	We construct a model for the spatial and velocity distributions of runaways.  
We use predicted ejection velocities from the supernova mechanism to eject stars from 
the exponential disk and then we track their orbits through throughout the Galaxy.  
In \S2, we describe the model and the results of our simulations.  We apply the simulations 
to the runaway HD~271791 in \S3, to HVS surveys in \S4, and to halo radial velocity surveys 
in \S5.  We conclude in \S6.

\section{The Simulations}

	The model we construct to explore the global velocity and spatial distribution 
of runaways in the Milky Way has two components: the gravitational potential of the 
Galaxy and the velocity distribution of ejected stars.  For the Galactic potential, 
we use the 3-component model defined in \citet{kenyon08}.  This disk, bulge, and halo 
model fits observations of the Galaxy on scales from 5 pc to 100 kpc, and has a circular 
orbital velocity of 220 \kms\ at $r=8$ kpc \citep[e.g.,][]{hogg05}.

	We consider runaways ejected from an exponential disk with a
representative distribution of ejection velocities, explore the propagation 
of runaways in the Galaxy using a suite of simulations, and examine the 
resulting distributions in Galactocentric radius $r$ and radial velocity 
\vrad\ at the end of the simulation. 

\subsection{Construction of the Simulations}

	To generate populations of runaways in the Galaxy, we perform Monte Carlo 
simulations of $10^6$--$10^7$ stars ejected into 3D orbits from the 
exponential disk.  Each star begins on a circular orbit with velocity
$\vec{v}_{0}(r_{init})$ at some distance \rinit\ from the Galactic center
in the plane of the Galaxy.  The star is ejected at an angle $\theta_i$
relative to the orbital velocity vector and an angle $b_i$ relative 
to the plane of the Galactic disk.  For a randomly oriented ejection velocity
$\vec{v}_{ej}$, the initial velocity of the ejected star is
$\vec{v}_{init} = \vec{v}_{0} + \vec{v}_{ej}$. For a star ejected in the
direction of Galactic rotation ($\theta_i \approx$ 0), \rinit\ becomes the 
pericenter of its orbit; for a star ejected opposite Galactic rotation 
($\theta_i \approx \pi$), \rinit\ becomes the apocenter of its orbit.

	To follow the trajectories of these ejected stars through the Galaxy, 
we integrate the equations of motion numerically. In \citet{bromley06} and 
\citet{kenyon08}, we used a simple leap-frog integrator to track stellar orbits 
through a 1D Galactic potential.  To provide more accurate solutions for the 
trajectories of runaways in a 3D Galactic potential, here we use an adaptive 
fourth-order integrator with Richardson extrapolation 
\citep[e.g., equations 1--3 of][see also Chapter 15 of Press et al. 1992]{bk06}.  
Starting at \rinit\ with velocity $\vec{v}_{init}$, the code integrates the 
full 3D orbit through the Galactic potential to track position and velocity as 
a function of time.  We integrate the orbit for a random time $t$, where $t$ lies 
between 0 and $t_{ms}$, the main-sequence lifetime of the runaway star.  For stars 
ejected during the supernova explosion of a more massive and much shorter-lived 
companion star, $t$ is roughly the age of the runaway star.

	To motivate choices for the initial conditions of our simulations, we derive
the approximate velocities of runaways capable of reaching the outer halo of the Galaxy.  
Numerical simulations of binary systems disrupted by supernova explosions show that 
runaways have maximum ejection velocities of $\sim$ 400 \kms\ \citep{portegies00,dray05}.  
Realistic dynamical ejection models yield similar maximum ejection velocities
\citep{leonard93,fregeau04}. If the highest velocity runaways are ejected in the
direction of Galactic rotation, they receive an additional kick of $\approx$ 200
\kms. Thus, the maximum possible ejection velocity is roughly $v_{ej,max} \approx$
600~\kms.  Unbound HVSs in the outer halo have $v_{rad} \gtrsim$ 400 \kms\ at $ r
\approx$ 70 kpc \citep{brown09a}.  To match these properties in our model for the
gravitational potential of the Milky Way,
runaways must have initial velocities \vej\ = 725~\kms\ at \rinit = 1~kpc, \vej\ =
550~\kms\ at \rinit = 10~kpc, and \vej\ = 500~\kms\ at \rinit = 30~kpc \citep[see
Fig. 2 of][]{kenyon08}.  Thus, runaways ejected from \rinit\ $\lesssim$ 5~kpc cannot
reach the outer halo. This conclusion leads us to consider a fiducial set of
simulations with \vej = 400 \kms\ and \rinit\ = 10 kpc.

	Long travel times through the halo constrain our choices for the initial ages and 
masses of main sequence runaway stars. Stars with \vej\ = 400 \kms\ travel $\sim$ 40 kpc 
in $\sim$ 100~Myr. Thus, stars with main sequence lifetimes $t_{ms}$ $\gtrsim$ 100~Myr and
initial masses $m \lesssim$ 4.5~\msun\ can reach the outer halo 
\citep{schaetal92,schaetal93,dem04}.  Stars with $m \approx$ 4--6~\msun\ and 
$t_{ms} \approx$ 65--160 Myr can reach the inner halo.  To avoid confusion with 
indigenous halo stars \citep[see][]{brown06, brown09a}, we focus on main sequence 
stars with $m$ = 1.5--6~\msun\ and $t_{ms}$ = 65~Myr to 2.9 Gyr \citep[][see also 
Table 1]{schaetal92,schaetal93,dem04}. These lifetimes are much longer than the 
lifetimes, $t_{ms} \lesssim$ 10--20~Myr, of their $m \gtrsim$ 10~\msun\ binary 
companions that explode as supernovae \citep{schaetal92,schaetal93}.  Thus, we 
assume for simplicity that runaways have ages of zero when they are ejected from 
the binary \citep[see also][]{portegies00}. 

\subsection{Results for a Fiducial Simulation}

\begin{figure}
\includegraphics[width=6.5in]{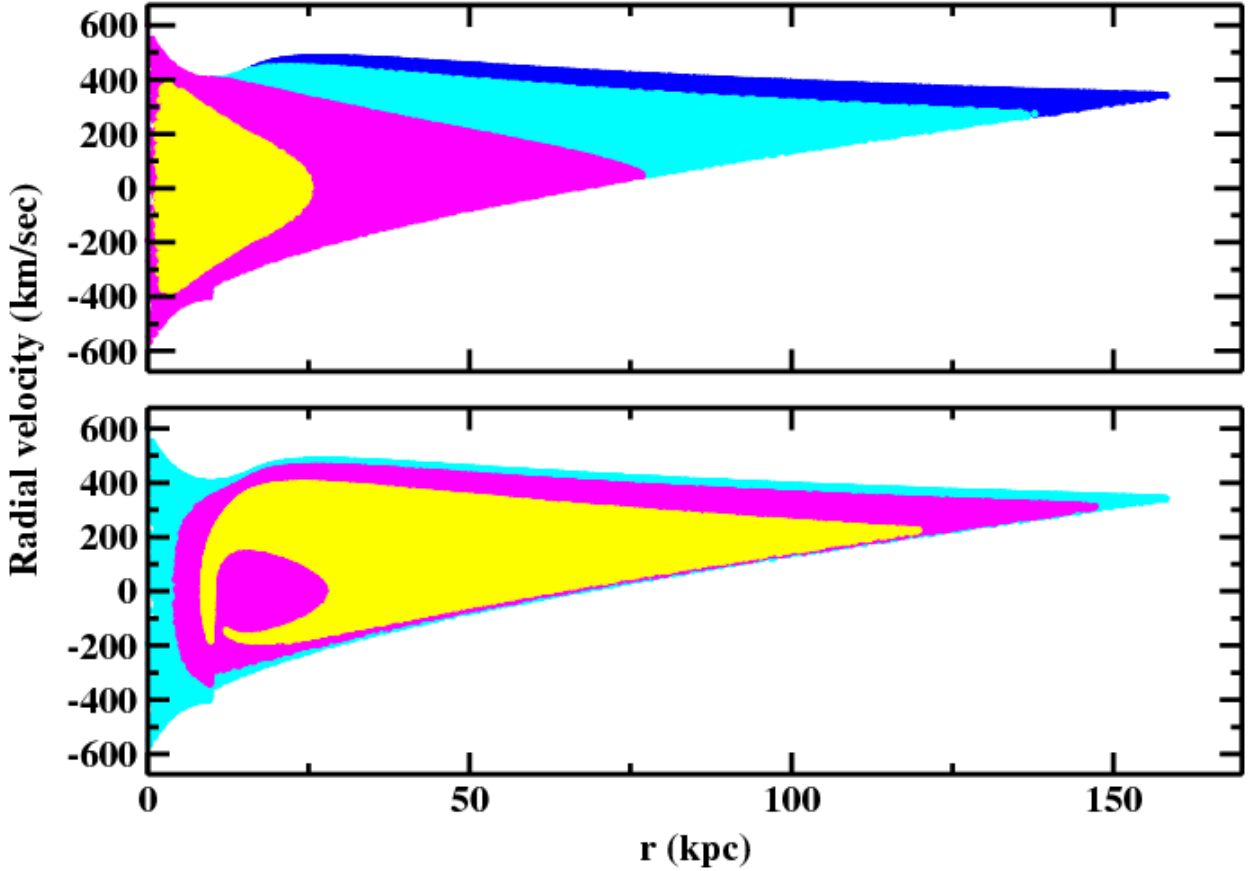}
\vskip 2ex
\caption{\label{fig:rmv2}
Scatter plots of simulated runaways ejected with \rinit\ = 10~kpc, \vej\ = 400~\kms, 
and $m$ = 3~\msun.  Each point is a ($r$,\vrad) pair derived from a Monte Carlo 
simulation of $10^6$ stars. 
{\it Upper panel:} points ejected into the Galactic disk ($b_i <$ 30\deg) are 
color-coded according to the ejection angle $\theta_i$ relative to the direction 
of Galactic rotation.  Blue (yellow) points indicate stars ejected with 
$\theta_i = -\pi / 4 \rightarrow \pi /4$  
($\theta_i = 3 \pi / 4 \rightarrow 5 \pi /4$).  Cyan and magenta points indicate 
stars ejected at intermediate angles, $\theta_i = \pm (\pi / 4 \rightarrow \pi / 2)$ 
for cyan points and $\theta_i = \pm (\pi / 2 \rightarrow 3 \pi / 4)$ for magenta points.
Stars ejected along (against) the direction of Galactic rotation reach the largest
(smallest) distances in the Galactic halo.
{\it Lower panel:} points are color-coded according to $b_i$, the ejection angle
relative to the Galactic plane, with $b_i <$ 30\deg\ in cyan, 30\deg\ $ < b_i <$ 
50\deg\ in magenta, and $b_i >$ 50\deg\ in yellow.  Stars ejected into the plane
(halo) reach the largest (smallest) distances from the Galactic center.
}
\end{figure}

	We first investigate the Galactic distribution of runaways for the fiducial 
parameters \rinit\ = 10 kpc, \vej\ = 400 \kms\, and $m = 3$ \Msolar.  
Figure~\ref{fig:rmv2} shows results for $10^6$ runaways; we plot \vrad\ as a function 
of $r$. In the upper panel, we color-code the points in four intervals of $\theta_i$ 
for stars with $b_i <$ 30\deg.  Blue (yellow) points indicate stars ejected with 
$\theta_i = -\pi / 4 \rightarrow \pi /4$  ($\theta_i = 3 \pi / 4 \rightarrow 5 \pi /4$).
Cyan and magenta points indicate stars ejected at intermediate angles, 
$\theta_i = \pm (\pi / 4 \rightarrow \pi / 2)$ for cyan points and 
$\theta_i = \pm (\pi / 2 \rightarrow 3 \pi / 4)$ for magenta points.  
In the lower panel, we color-code the points in three intervals of $b_i$: 
$b_i < $ 30\deg (cyan), 30\deg\ $< b_i < $ 50\deg (magenta), and $b_i > $ 50\deg (yellow). 

	The distribution of runaway stars in Fig. \ref{fig:rmv2} has several 
features.  In the top panel, stars with the largest initial velocities (blue points) 
are marginally bound to the Galaxy. Thus, they travel far into the halo, reach a maximum
distance $r_{max}$, and then fall back towards the Galactic Center. However, the
travel time for this orbit, $t \approx \pi ~ r_{max}$ / \vinit\ $\sim$ 1 Gyr,
is longer than the main sequence lifetime of a 3 \msun\ star, $t_{ms} \approx$ 
350~Myr. These stars can only be observed on their outward path through the halo and 
always have positive radial velocities. Thus, the sequence of blue points from 
($r$, \vrad) = (20, 500) to (160, 340) is an age sequence, with 20--30 Myr-old stars 
at small $r$ and 300--350 Myr-old stars at the largest $r$.  

	Stars ejected opposite to the direction of Galactic rotation (yellow points) 
have the smallest initial velocities and thus cannot reach large $r$.  These stars live 
long enough to orbit the Galaxy at least once. Thus, the group of yellow points represents 
a mixture of young and old stars with a velocity distribution symmetric about zero.

	Stars ejected at intermediate angles (cyan and magenta points) fill the 
($r$, \vrad) space in between the group ejected along or against Galactic rotation. 
For each group, the sequence from largest \vrad\ to largest $r$ is an age sequence, 
with younger stars at small $r$ and older stars at large $r$. Thus, most of the lower 
envelope of the complete ensemble of points, extending from ($r$, \vrad) = (30, $-$200)
to ($r$, \vrad) = (160, 340) consists of old stars near the end of their main sequence 
lifetimes.

	The lower panel of Fig. \ref{fig:rmv2} illustrates how the maximum radial
velocities decrease with increasing $b_i$.  Stars at high latitude (yellow points;  
$b_i \ge 50^{\circ}$) have the smallest velocities, with \vrad\ $\lesssim$ 400 \kms\
at all $r$. Stars at intermediate latitudes (magenta points;  $30^{\circ}$ $\le b_i
\le$ $50^{\circ}$) reach \vrad\ $\lesssim$ 450 \kms\ at all $r$;  stars at low
latitudes (cyan points; $b_i \le$ $30^{\circ}$) have the largest final velocities
(\vrad\ $\sim$ 500 \kms) and reach the largest final radii, $r$ = 150--160~kpc in
350 Myr.  The form of the Galactic potential produces this variation. Stars ejected
perpendicular to the Galactic disk receive a smaller kick from Galactic rotation
compared to stars ejected parallel to the disk. Once stars leave the plane of the
disk, they also feel the full disk potential. Thus, stars ejected perpendicular to
the disk have smaller initial radial velocities and decelerate faster than stars
ejected parallel to the disk.

To show how the results depend on each input parameter, Figure~\ref{fig:rmv4} plots the
velocity distributions for simulations where we vary each of the three fiducial parameters 
separately and hold the other two fixed. Here, we color code the points with their final
galactic latitude.  Instead of illustrating relative densities, our intent in this figure 
is to show changes in the shape of the distributions of \vrad\ and $r$ as functions of 
the various input parameters and $b_f$, the Galactic latitude of stars at the end of 
the simulation. The upper left panel of Figure~\ref{fig:rmv4} repeats the distribution 
of points in the lower panel of Fig. \ref{fig:rmv2} with a different color coding. Here, 
the variation of \vrad\ and $r_{max}$ with final latitude is much more pronounced than 
the variation with ejection angle.

	The predicted distributions clearly depend on \rinit\ and \vinit\ (upper right and 
lower left panels of Figure~\ref{fig:rmv4}). Stars ejected at small $r$ have to climb out 
of a deeper potential well than stars ejected at large $r$. At fixed \vej, runaways ejected 
from 30 kpc have larger velocities in the outer halo (maximum \vrad\ = 500 \kms) than 
stars ejected from 10 kpc (maximum \vrad\ = 400 \kms). Similarly, stars with smaller 
(larger) ejection velocities maintain smaller (larger) velocities throughout their 
passage through the Galaxy. Thus, at fixed \rinit, runaways ejected at larger velocities 
have larger velocities in the outer halo and reach larger distances in the halo.

	Finally, stellar lifetimes have a significant impact on the distributions
of position and velocity (Figure~\ref{fig:rmv4}; lower right panel). For fixed
initial ejection velocities, stars with longer main sequence lifetimes travel 
farther into the halo. Stars with longer lifetimes can reach large distances with 
smaller initial radial velocities.  Thus, lower mass stars have smaller \vrad\ at 
larger $r$. For our adopted main sequence lifetimes, 1.5--4 \msun\ runaway stars 
ejected from 10 kpc have asymptotic radial velocities
\begin{equation}
v_{\rm rad} (r_{max,10}) \approx 250 ~ (M_{\star} / 2~M_{\odot})^{2/3} ~ {\rm km~s^{-1}}
\end{equation}
at maximum Galactocentric distances
\begin{equation}
r_{max,10} \approx 400 ~ (M_{\star} / 2~M_{\odot})^{-2.4} ~ {\rm kpc} ~ .
\end{equation}
At 30 kpc, runaways have larger asymptotic radial velocities
\begin{equation}
v_{\rm rad} (r_{max,30}) \approx 400 ~ (M_{\star} / 2~M_{\odot})^{1/6} ~ {\rm km~s^{-1}}
\end{equation}
at maximum Galactocentric distances
\begin{equation}
r_{max,30} \approx 575 ~ (M_{\star} / 2~M_{\odot})^{-2.6} ~ {\rm kpc} ~ .
\end{equation}
In both equations for $r_{max}$, the large exponent follows from the mass dependence
of the main sequence lifetime \citep[$t_{ms} \propto M_{\star}^{-3}$;][]{schaetal92,
schaetal93,dem04}. 

Tables \ref{tab:fidmed}--\ref{tab:fiddisp} quantify these generic conclusions for an 
expanded set of fiducial simulations.  For simulations with \rinit\ = 10 kpc, the median 
velocities in Table \ref{tab:fidmed} show clear trends with stellar mass and $r$. 
Only the highest velocity runaways can reach large $r$; for all stellar masses, the 
median velocity increases with increasing $r$.  Because their main sequence lifetimes
are longer, lower mass stars to orbit the Galactic Center 2--3 times. Thus, median 
velocities for lower mass stars are closer to zero than those for higher mass stars. 

The trends of median radial velocity with $b_f$ depend on $r$ and stellar mass. 
Stars ejected into the disk have larger initial velocities and always travel farther 
than stars ejected perpendicular to the disk. Thus, at large $r$, stars at smaller
galactic latitude have larger median \vrad.  Higher mass stars show the largest variation 
of median \vrad.  For every 30$^\circ$ increase in $b_f$, the median \vrad\ declines 
$\sim$ 25--75 \kms\ for 1.5--3 \msun\ stars.  At small $r$, 1.5--3~\msun\ stars with 
$b_f <$ 30\deg\ and $b_f >$ 50\deg\ were ejected against Galactic rotation; these stars 
have a range of orbital phases and thus have median \vrad\ close to zero. At intermediate
$b_f$, there is a mixture of stars ejected opposite to Galactic rotation (which have
median \vrad\ close to zero) and stars ejected with Galactic rotation (which have larger
median \vrad). Thus, the median velocities for $b_f$ = 30\deg--50\deg\ are generally 
large.

The typical radial velocity dispersions of runaway stars also vary consistently with 
$r$, $b_f$, and stellar mass (Table \ref{tab:fiddisp}). Stars observed at small $r$ 
are a mix of young stars ejected into the outer Galaxy and older stars orbiting in 
the inner Galaxy. Thus, these stars have large velocity dispersions for all stellar 
masses.  Because massive stars have shorter lifetimes, they do not travel far from 
their ejection point and tend to have smaller velocity dispersions. Stars observed at 
large $r$ are older stars traveling on extended bound orbits. None of these stars live
long enough to fall back into the Galaxy. Thus, all have large, positive \vrad\ and
small velocity dispersions.

\subsection{Results for a Galactic Distribution of Runaways}

	We now consider a simulation for a complete ensemble of runaways ejected 
from the full Galactic disk. To make this simulation, we adopt probability
functions to assign \rinit\ and \vej\ for each runaway star, integrate the orbit 
through the Galaxy, and derive the radial velocity \vrad\ and position $r$ for 
each runaway at a random time $t$ in its orbit.  Our approach differs from 
\citet{davies02}, who simulated runaways ejected uniformly from $1 < r < 10$ kpc. 

	For the initial distribution of \rinit, we assume stars are ejected from
an exponential disk with a radial scale length of 2.4 kpc \citep{siegel02}. We
adopt an inner radius of 3~kpc and an outer radius of 30~kpc \citep{brand2007}.  
Thus, the probability distribution for \rinit\ is 
\begin{equation}
\label{eq:pdfr}
p(r_{\rm init}) \propto r_{init}~e^{-r_{\rm init}/2.4~{\rm kpc}}
\end{equation}
Runaways ejected from \rinit\ $<$ 3 kpc cannot reach the outer halo. Runaways 
ejected from \rinit\ $>$ 30 kpc are too rare to make a significant impact on the 
derived distribution of ($r$, \vrad). 

	For the initial distribution of \vej, we adopt results from published
analyses of runaways ejected from binary systems disrupted by supernovae.  Although 
dynamical encounters can also produce runaways, there are no published simulations 
predicting $p(v_{\rm ej})$ 
the probability distribution of ejection velocities.  For \vej\ $\sim$ 
20--400 \kms, a simple function,
\begin{equation}\label{eq:pdfv}
p(v_{\rm ej}) \propto e^{-v_{\rm ej}/150~{\rm km~s^{-1}}},
\end{equation} 
provides a reasonable match to the \citet{portegies00} simulations of 
binary supernova ejections.  

We select the \citet{portegies00}
ejection velocity distribution function because it is physically well
motivated and because a similar distribution function is not available for
the dynamical ejection mechanism.  Dynamical ejections can attain higher
velocities \citep{leonard88, leonard90, leonard91, leonard93}; however,
the theoretical maximum ejection velocity is not realizable because
compact binary interactions are more likely to merge stars than to eject
runaways \citep{fregeau04}.  Moreover, the ejection rate from binary-binary 
encounters is probably smaller than supernova ejections for intermediate 
mass stars. Dynamical ejections depend on the joint probability of colliding 
two binaries within the main sequence lifetime of the stars \citep[][]{brown09a}.  
Thus we use the binary-supernova ejection velocity distribution as representative 
of the runaway process for 1.5--6 \Msolar\ stars.

\begin{figure}
\includegraphics[width=6.5in]{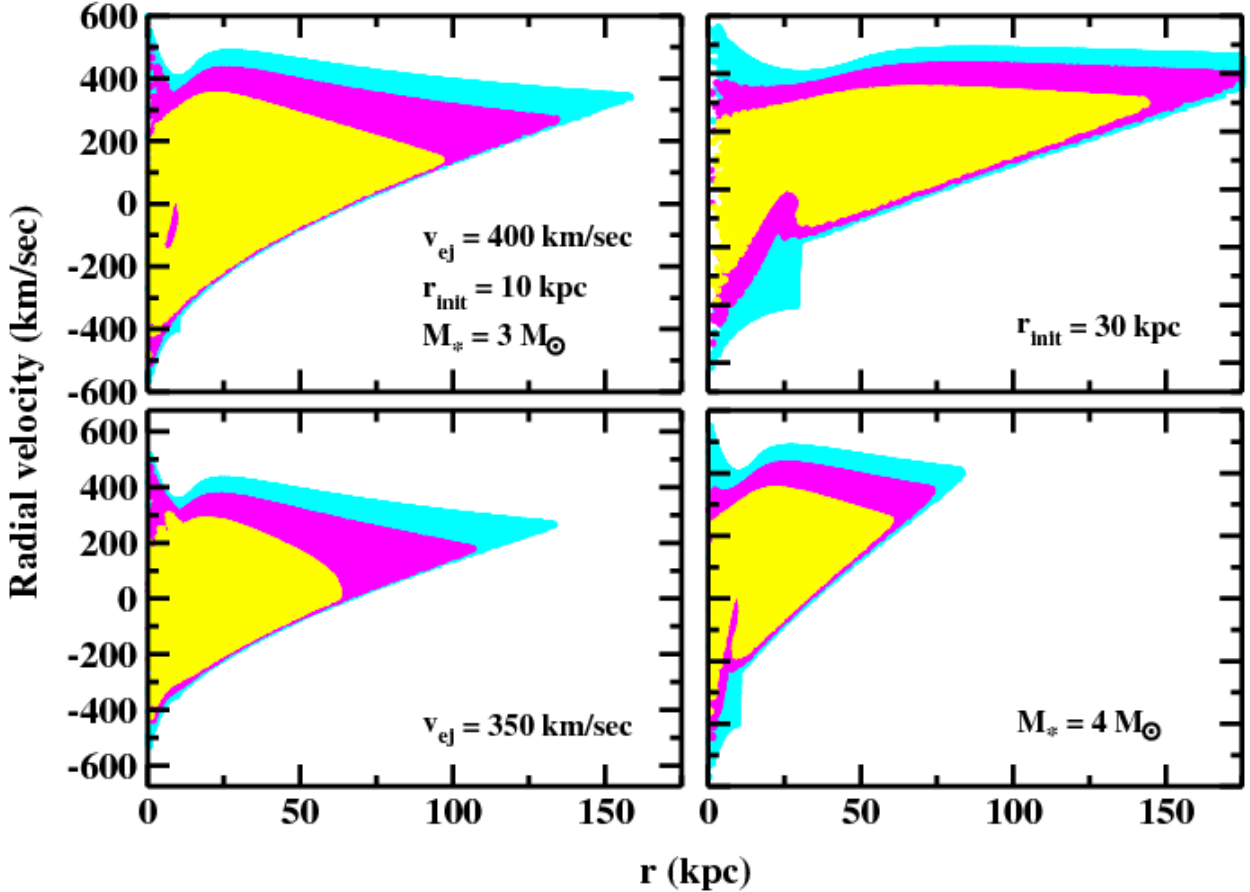}
\vskip 2ex
\caption{\label{fig:rmv4}
Scatter plots of simulated runaways ejected with fixed initial radius
\rinit, ejection velocity \vej, and stellar mass $m$.  Each point is a
($r$,\vrad) pair derived from a Monte Carlo simulation of $10^6$ stars.
In each panel, the colors of the points indicate the Galactic latitude
at the end of the simulation; cyan: $b \le 30^\circ$, magenta: 
$30^\circ < b \le 50^\circ$, and yellow: $b > 50^\circ$.  
{\it Upper left panel:} results of the fiducial simulation with \rinit\ =
10 kpc, \vej = 400 \kms, and $m$ = 3 \msun.  Runaways ejected parallel to 
the plane of the Galactic disk achieve higher \vrad\ at larger $r$ than 
runaways ejected perpendicular to the disk plane. 
{\it Upper right panel:} results for \rinit\ = 30 kpc. Runaways ejected
at larger \rinit\ achieve larger \vrad\ at larger $r$.
{\it Lower left panel:} results for \vej = 350 \kms. Runaways ejected
with smaller \vej\ have smaller \vrad\ at all $r$.
{\it Lower right panel:} results for $m$ = 4 \msun. Smaller main sequence
lifetimes do not allow more massive stars to reach large $r$.
}
\end{figure}

\begin{figure}
\includegraphics[width=6.5in]{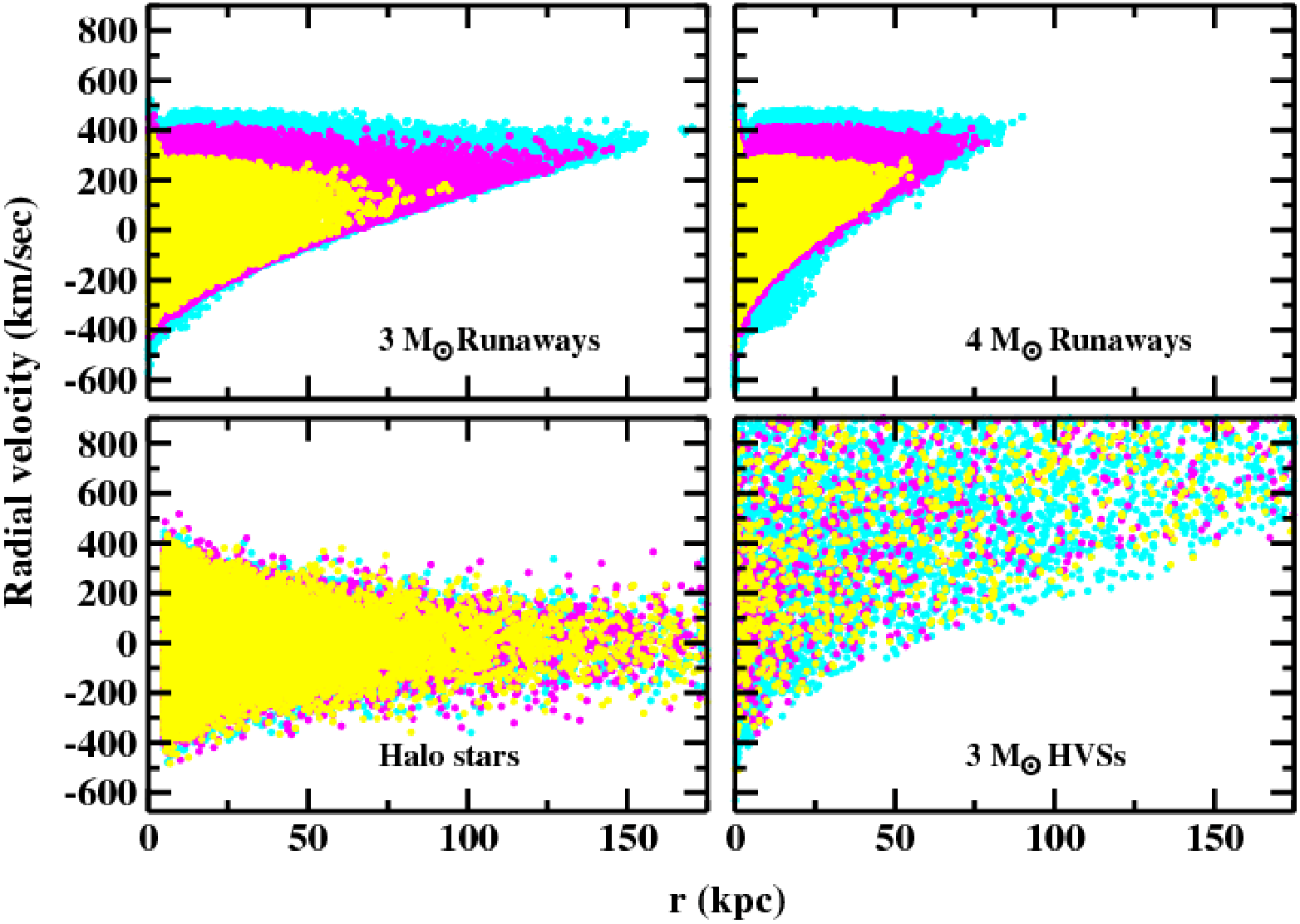}
\vskip 2ex
\caption{\label{fig:rmv4-h}
Comparison of ($r$, \vrad) diagrams for halo stars, HVSs, and runaways.
Stars are color-coded according to their Galactic latitude;
cyan: $b_f <$ 30\deg, magenta: 30\deg $< b_f <$ 50\deg, and
yellow: $b_f >$ 50\deg.  The top panels shows results for 3 \msun\ (left 
panel) and 4 \msun\ (right panel) runaways ejected from the Galactic disk 
with probability distributions for \vej\ (Eq. \ref{eq:pdfv}) and for 
\rinit\ (Eq.  \ref{eq:pdfr}). Longer-lived lower mass runaway stars reach 
larger $r$ than shorter-lived more massive runaways. Runaways ejected into 
the disk achieve larger $r$ than stars ejected perpendicular to the disk.
The lower left panel plots the expected velocity distribution of halo
stars with a velocity dispersion of 110 \kms\ and a $n \propto r^{-3}$
density law. Although nearby halo stars can have large \vrad, distant
runaways have much larger \vrad\ than distant halo stars.
The lower right panel shows the expected velocity distribution of HVSs
ejected from the Galactic Center. At each $r$, the lower limit to \vrad\ for
HVSs follows the predicted lower limit for runaway stars. However, at all
$r$ and $b_f$, HVS have much larger \vrad\ than runaway stars.
}
\end{figure}

Full-disk simulations of runaways yield many of the same features in the \vrad-$r$ 
diagram (Figure~\ref{fig:rmv4-h}, top panels).  As in Figure~\ref{fig:rmv4}, the lower 
envelope of the set of points is defined by an ensemble of old runaways close to the end
of their main sequence lifetimes. Stars ejected into the disk (cyan points) receive 
the maximum ejection velocity. Thus, these stars have the largest \vrad\ at all $r$.
Stars ejected into the halo (yellow points) have the smallest ejection velocity,
the smallest \vrad\ at all $r$, and the smallest radial extent.  More massive stars 
with shorter $t_{ms}$ do not live long enough to reach large $r$.  Thus, lower mass 
stars have more extended radial distributions than more massive stars.

The major difference between the fiducial and the full-disk simulations is the variation 
of the maximum \vrad\ with $r$.  Runaways ejected from the inner disk slowly 
decelerate as they move from their point of origin. In an ensemble of
runaways ejected from a single radius in the inner disk, stars at larger $r$ 
therefore have smaller \vrad\ than stars at smaller $r$ (upper left panel of 
Figure~\ref{fig:rmv4}). However, runaways ejected from the outer disk coast 
outward at roughly constant \vrad\ (upper right panel of Figure~\ref{fig:rmv4}). 
These two features of the evolution combine to produce a roughly constant maximum 
\vrad\ with $r$ in the top panels of Figure~\ref{fig:rmv4-h}. This maximum 
velocity is independent of stellar mass and is roughly 100 \kms\ larger than the 
maximum ejection velocity.

Tables \ref{tab:fullmed}--\ref{tab:fullvdisp} list the median velocities, radial 
velocity dispersions, and vertical velocity dispersions for the full-disk simulations of 
1.5--6 \msun\ stars. For all stellar masses, stars ejected at larger \vrad\ reach 
larger $r$. Thus, the median radial velocity increases with $r$. Because the most
distant stars must have roughly the same high ejection velocity to reach large $r$,
these stars have smaller velocity dispersions than the mix of low and high velocity
runaways at small $r$.  For stars ejected with similar velocities, lower mass stars 
live longer and can reach larger $r$. Thus, the median \vrad\ and the velocity 
dispersions increase with increasing \mstar.  

The distribution of runaways in the \vrad-$r$ diagram differs from the distribution
of HVSs (Figure~\ref{fig:rmv4-h}, lower right panel). The lower envelope of the HVS
distribution is identical to the runaway distribution and is composed of stars
with ages comparable to their main sequence lifetimes. However, the distribution
of HVSs has three features not observed in runaway stars. The HVS distribution has a 
core of stars at $r \lesssim$ 3 kpc; these stars were ejected from the \GC\ at small velocities 
($\lesssim$ 700 \kms) and cannot reach large $r$ \citep{kenyon08}. Because some
HVSs are ejected at very high speeds ($\gtrsim$ 1200 \kms), these stars can reach 
large $r$ with radial velocities much larger than any runaway star. Finally,
the velocities of HVSs ejected isotropically from the \GC\ are independent of
$b$. Thus, high velocity HVSs are observable at all $b$. In contrast, high velocity
runaway stars are only observable at low galactic latitude.

The velocity distribution of runaways is also very different from the velocities of
halo stars (Figure~\ref{fig:rmv4-h}, lower left panel). Halo stars have a 1D
velocity dispersion of roughly 100--110 \kms\ \citep{helmi08,brown09a} and a spatial
density close to $n \propto r^{-3}$ \citep{siegel02}. Thus, the envelope of the halo
velocity distribution is symmetric about zero and declines slowly with radius. At
large $r$, there are many halo stars with \vrad\ $\approx$ 0. In contrast, there are
no intermediate mass HVSs or runaways with \vrad\ $\approx$ 0 at large $r$.

\begin{figure}
\includegraphics[width=5.5in]{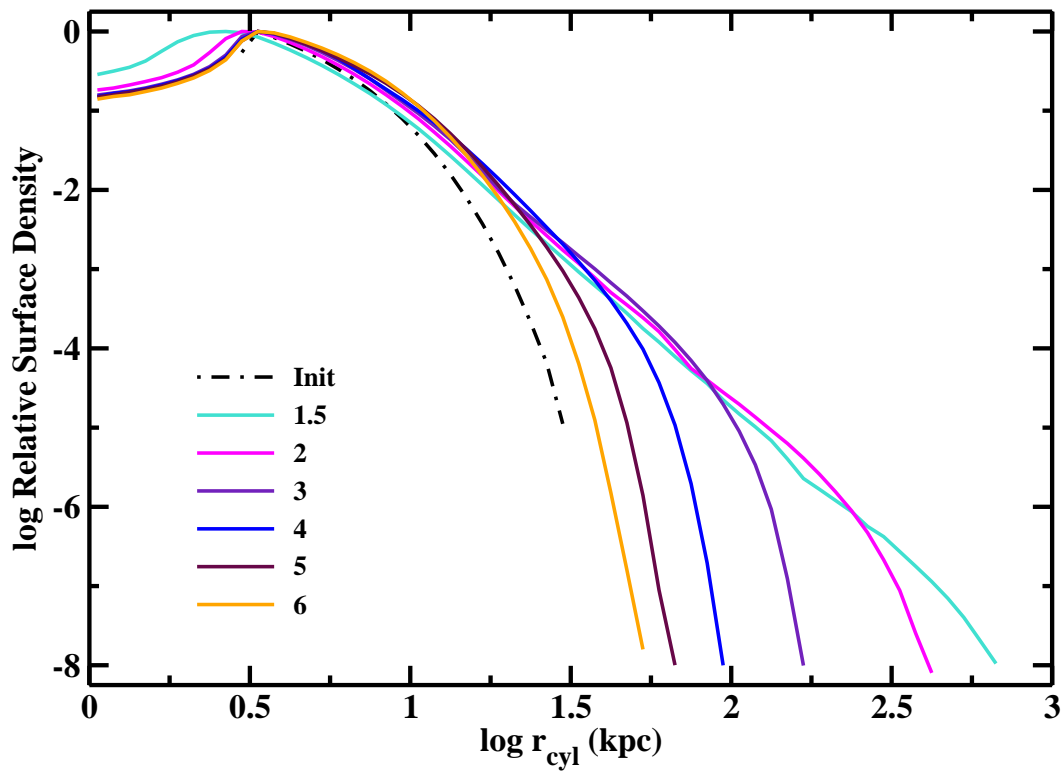}
\vskip 4ex
\caption{\label{fig:rad-den}
Predicted surface density distributions as a function of cylindrical radius
($r_{cyl}^2 = x_f^2 + y_f^2$) for runaways ejected with a range of velocities
(Eq. \ref{eq:pdfv}) from an exponential disk (Eq. \ref{eq:pdfr}). The dot-dashed black
line plots the initial density distribution. The colored lines plot surface density 
distributions for stars of different masses, as indicated in the legend. For all stars,
the density rises slowly from 1--3 kpc, follows a power law ($\Sigma \propto r^{-3.6}$)
at intermediate radii, and then falls exponentially at large radii.
}
\end{figure}

To quantify the differences between runaways, HVSs, and halo stars, we now consider 
the predicted radial surface density of runaways (Figure~\ref{fig:rad-den}).  Compared 
to the initial surface density of the Galactic disk (dot-dashed line in 
Figure~\ref{fig:rad-den}), runaways are much more radially extended. Runaways ejected against
Galactic rotation populate the inner disk ($r \lesssim$ 3 kpc). These runaways have
a large velocity dispersion at all $b$ (top panels of Figure~\ref{fig:rmv4-h}).  
Runaways ejected at high velocity along Galactic rotation populate the outer disk.
These ejections produce a power-law surface density profile, $\Sigma \propto r^{-n}$ 
with $n$ = 3.5--3.6, at intermediate $r$ and an exponential decline at large $r$.
Longer-lived lower mass stars have the most extended power-law component. Short-lived
massive stars have exponential density profiles more similar to the density profile of
the Galactic disk.

Although the density profiles of 1.5--3 \msun\ runaways are extended, they are much steeper 
than the density profiles of halo stars and HVSs. Fits to observations of halo stars 
typically yield power law density profiles with $n \approx$ 2.7--3.5 
\citep[][and references therein]{helmi08}.  Our numerical simulations suggest that 
low mass HVSs have a bound component with $n$ = 3 and an unbound component with 
$n$ = 2--2.5 \citep[][see also Hills 1988]{kenyon08}.  The short, finite lifetimes 
of massive stars steepens the HVS density profile ($n \gtrsim$ 3) at $r \gtrsim$ 80 kpc.  
However, these density profiles remain shallower than the density profiles of 
runaway stars.

\begin{figure}
\includegraphics[width=6.5in]{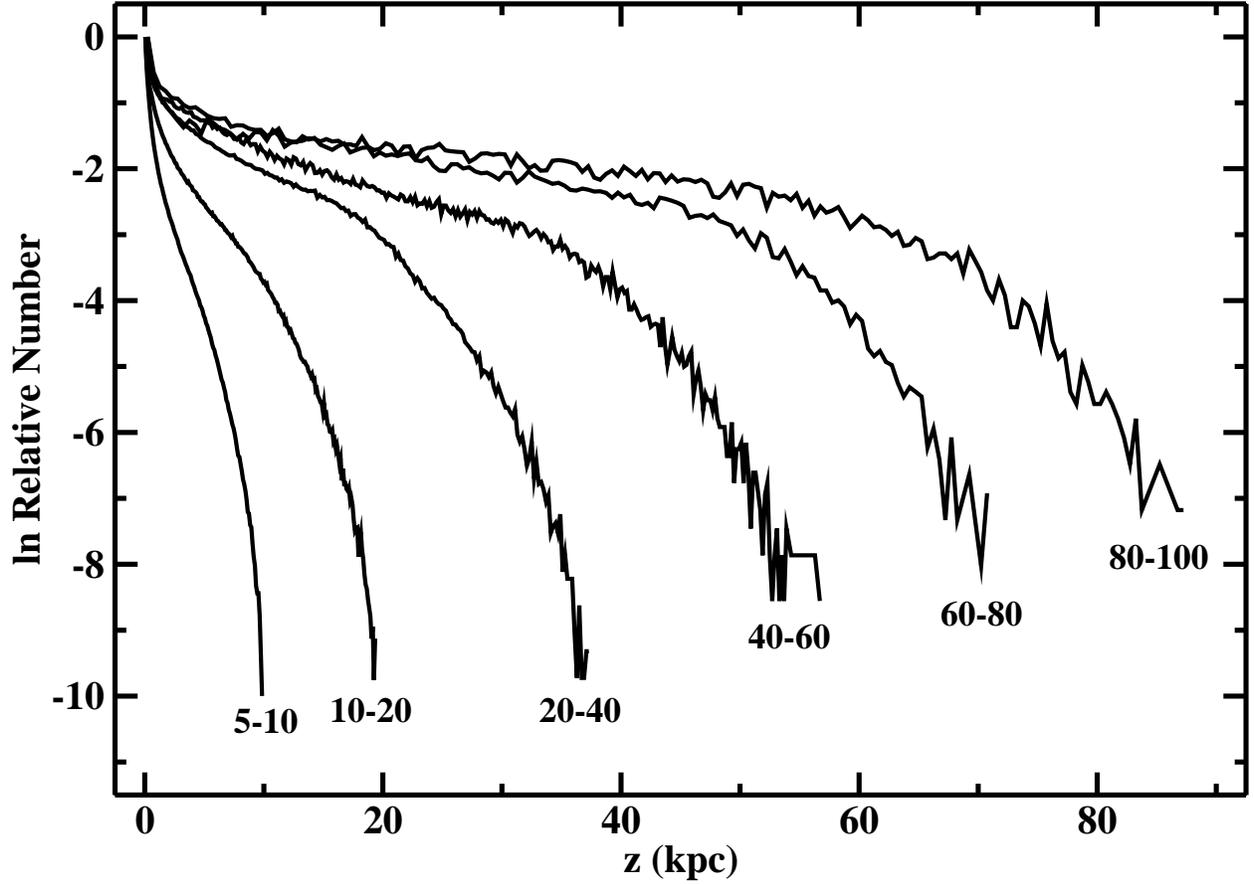}
\vskip 3ex
\caption{\label{fig:z-den}
Predicted natural log of the relative number density as a function of $z$,
the height above the disk plane, for 2 \msun\ runaway stars. The curves plot the 
number density in six cylindrical radius bins, as indicated.  
In each bin, the density distribution has a central core with a scale height 
of 300--1000 pc and an extended halo with a scale height of 2--30 kpc. The $z$
scale height of the extended halo increases with increasing cylindrical radius.
}
\end{figure}

The predicted vertical density distribution also distinguishes runaway stars from halo 
stars and HVSs (Figure~\ref{fig:z-den}). This density distribution has two components.  
Low velocity runaways produce a `thick disk' with a vertical scale height of 300--1000 pc.  
Although more distant runaways have slightly larger `thick disk' scale heights, the 
scale height is independent of stellar mass.  High velocity runaways lie in an 
extended disk-shaped halo with a vertical scale height of 2--40 kpc.  More distant 
runaways have much larger vertical scale heights. At $r \lesssim$ 50--60 kpc, the scale 
height of the extended halo is independent of stellar mass (Table \ref{tab:fullzheight}).  
At $r \gtrsim$ 60 kpc, the vertical scale height depends on stellar mass.  Massive stars 
(\mstar\ $\gtrsim$ 4 \msun) will short main sequence lifetimes cannot reach large $r$ and 
thus have no measurable scale height. For stars with longer stellar lifetimes, lower mass 
stars have smaller scale heights at large $r$. These smaller scale heights result from low 
velocity ejected stars, which have time to reach large $r$ only for the lowest mass main 
sequence stars.

The disk-shaped density distribution of the highest velocity runaways differs from the 
spherically symmetric density distributions of halo stars and HVSs. In our simulations,
HVSs ejected from the \GC\ have a spherically symmetric, power law density 
distribution with $n \approx$ 2--2.5 \citep{kenyon08}. 
Halo stars are also distributed spherically symmetrically and have a steeper
radial density profile \citep[$n \approx$ 2.7--3.5;][]{helmi08}. Because HVSs have a 
shallower density profile than halo stars, it is easier to identify HVSs at large halo 
distances than at small halo distances \citep[e.g.,][]{brown05,bromley06,kenyon08}.
The steeper density profiles produced in our runaway star simulations suggest nearby 
runaways are easier to identify than distant runaways. We consider this possibility 
further in \S4--5.

\begin{figure}
\includegraphics[width=6.5in]{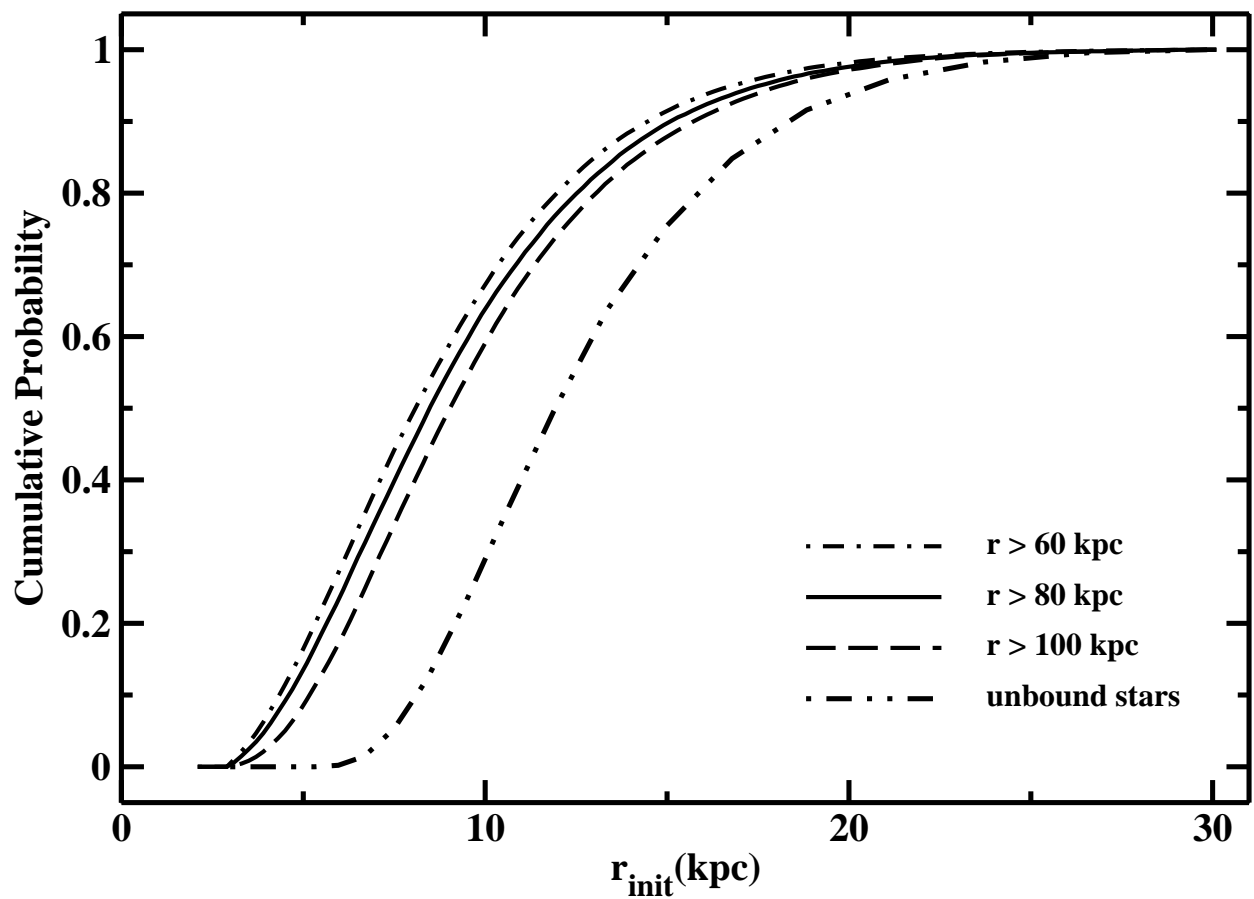}
\vskip 2ex
\caption{\label{fig:cum-prob}
Cumulative probability distributions as a function of initial disk radius for 
2 \msun\ runaway stars. Among runaways that reach final radii $r >$ 60 kpc (dashed 
line), $r >$ 80 kpc (solid line), and $r >$ 100 kpc (dot-dashed line), roughly 
50\% are ejected from inside the solar circle. Among {\it unbound} runaways
(double dot-dashed line), roughly 50\% are ejected from \rinit\ $\gtrsim$ 12 kpc.
}
\end{figure}

To conclude this section, we examine several additional properties of simulated runaways in 
the outer Galaxy.  Most runaways with $r \gtrsim$ 60 kpc are ejected from inside the solar 
circle (Figure~\ref{fig:cum-prob}).  Although most ejected stars have \rinit\ $\approx$
3--6 kpc, small ejection velocities prevent them from reaching $r \gtrsim$ 60 kpc (\S2.1). 
Because the depth of the Galactic potential is smaller for runaways with \rinit\ $\approx$ 
10--20 kpc, these stars make up a large percentage of stars with $r \gtrsim$ 60 kpc (see
also Figure~\ref{fig:rmv4}).  In addition, most runaways at large $r$ are low mass stars
with stellar lifetimes long enough to reach the outer Galaxy.  With median radial velocities 
of 10--200 \kms\ (Table \ref{tab:fullmed}), nearly all of these runaways are bound to the 
Galaxy.  Before reaching the outer galaxy, massive stars (\mstar\ $\gtrsim$ 4 \msun) evolve 
off the main sequence and are unobservable at $r \gtrsim$ 60 kpc. 

\begin{figure}
\includegraphics[width=6.5in]{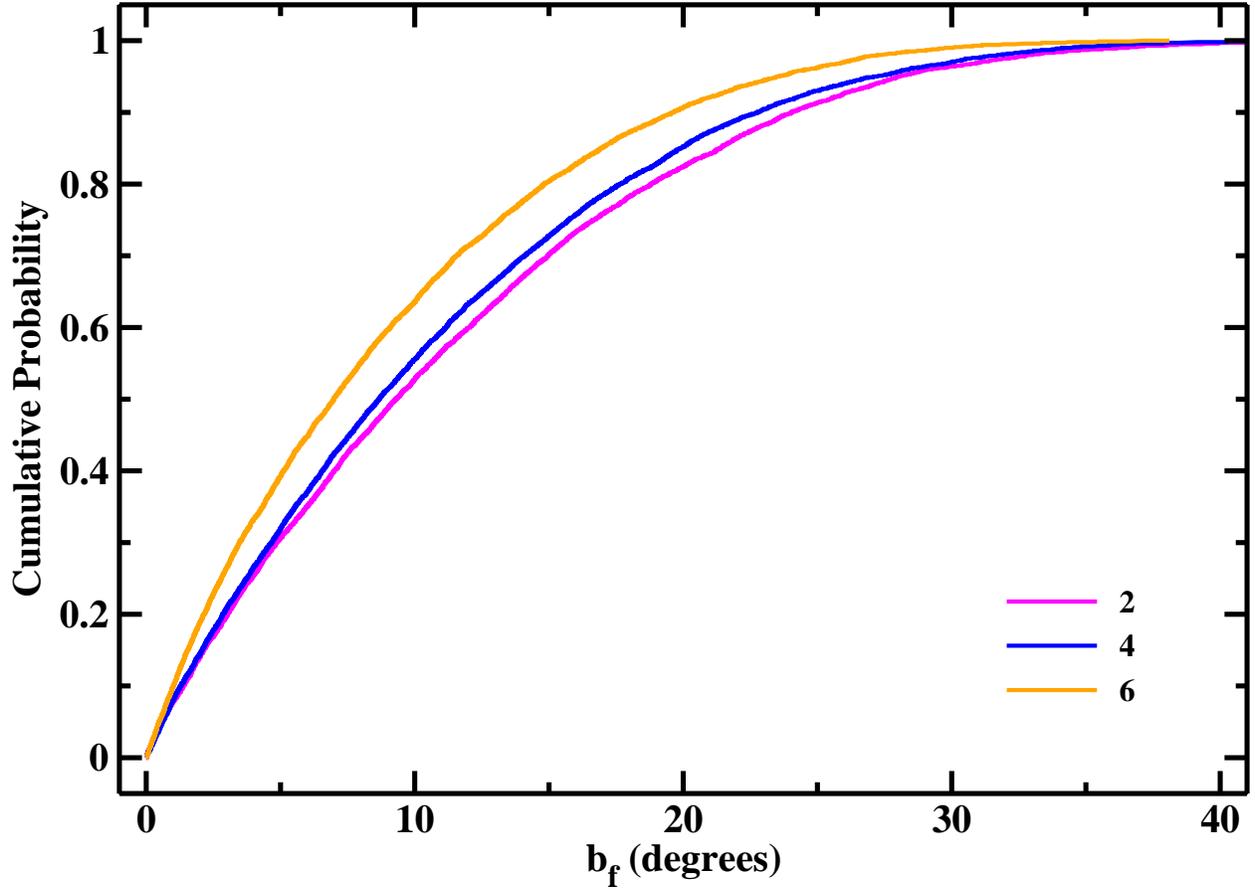}
\vskip 2ex
\caption{\label{fig:z-prob}
Cumulative probability distributions as a function of final galactic latitude for
unbound runaway stars. The legend indicates the stellar mass for each curve. The
median galactic latitude ranges from $b_{f,med}$ = 7\deg\ for 6 \msun\ stars to
$b_{f,med}$ = 9\deg--10\deg\ for 1.5--3 \msun\ stars. All unbound runaways have 
$b_f <$ 30 \deg.
}
\end{figure}

Our simulations yield a small fraction -- 0.07\% -- of runaways that are {\it not} bound to 
the Galaxy.  More than half of unbound runaways are ejected from outside the solar circle 
(double dot-dashed line in Figure \ref{fig:cum-prob}).  Nearly all unbound runaways have 
low Galactic latitude ($b_f <$ 30\deg; Figure \ref{fig:z-prob} and Table~\ref{tab:med-zr}).  
For 1.5--6 \msun\ stars, the shape of the cumulative probability function for $b_f$ is 
nearly independent of stellar mass. The median of the distribution, however, varies slowly 
with stellar mass. Our results suggest median Galactic latitude $b_{f,med}$ = 9\deg--10\deg\ for 
1.5--3 \msun\ stars and $b_{f,med}$ = 7\deg\ for 6 \msun\ stars (Table~\ref{tab:med-zr}).  

These results for the distribution of $b_f$ contrast with numerical simulations of HVSs, 
which yield a uniform distribution in $b_f$ \citep{bromley06,kenyon08}.
In the \citet{hills88} ejection mechanism, HVSs are ejected isotropically from the \GC.
For these stars, the disk is a small perturbation on the potential. Thus, simulated HVSs 
remain uniformly distributed in $b_f$ throughout their path through the Galaxy. For runaway
stars, Galactic rotation provides a significant kick to the ejection velocity. Stars
ejected with this rotation are more likely to have unbound velocities than other stars
(Figure \ref{fig:rmv2}--\ref{fig:rmv4-h}).  Thus, unbound runaways are confined to 
the plane of the disk.

\begin{figure}
\includegraphics[width=6.5in]{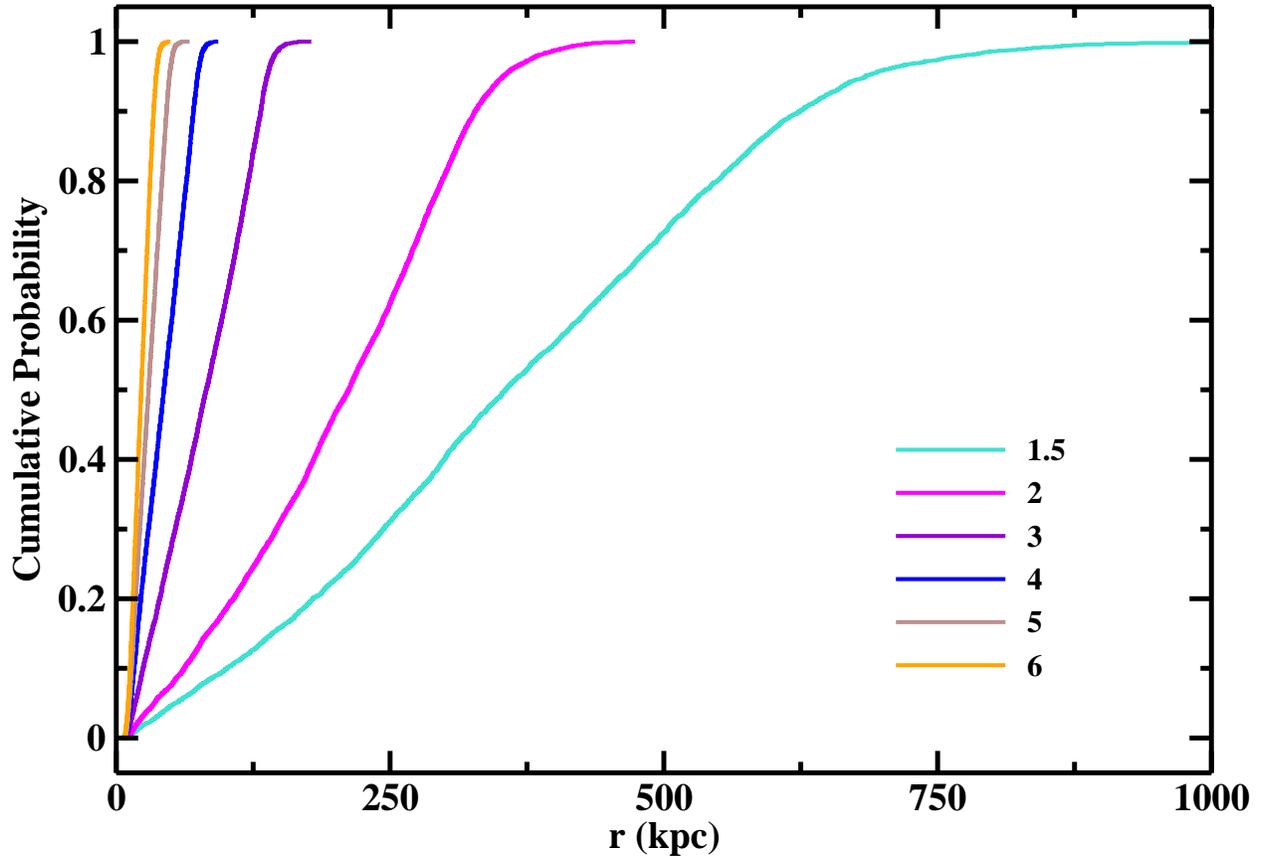}
\vskip 4ex
\caption{\label{fig:rf-prob}
Cumulative probability distributions as a function of distance for unbound runaway 
stars. The legend indicates the stellar mass for each curve. Lower mass runaways
reach larger $r$ than more massive runaway stars.
}
\end{figure}

Although unbound runaways have low $b_f$ independent of \mstar, the predicted distances 
are sensitive to \mstar\ (Figure~\ref{fig:rf-prob} and Table~\ref{tab:med-zr}).  The 
results in Figure~\ref{fig:rf-prob} show that the cumulative probability of $r$ grows 
roughly linearly from $r \approx$ 6 kpc to $r \approx$ 0.75$r_{max}$ and then asymptotically
approaches 1 at $r \approx r_{max}$. The limiting distance $r_{max}$ of unbound runaways 
is a strong function of stellar mass.  Long-lived low mass stars travel far into the 
outer Galaxy ($r_{max} \approx$ 500--1000 kpc for \mstar = 1.5--2 \msun); short-lived 
high mass stars rapidly evolve off the main-sequence and cannot reach large $r$ 
($r_{max} \approx$ 50--70 kpc for \mstar\ = 5--6 \msun). 

\subsection{Summary of the Simulations}

Our simulations demonstrate how stellar evolution and the Galactic potential combine
to influence the dynamical properties of runaway stars ejected from the Galactic disk.
Runaways that receive the maximum kick from the binary-supernova mechanism, $\approx$ 
400 \kms, can travel from the disk into the halo. These stars produce an extended 
disk-shaped distribution of stars, where the radial and vertical scale lengths are 
much larger than those of the main stellar disk. The size of this extended disk is very 
sensitive to stellar mass. Massive stars with short stellar lifetimes are much less extended 
than long-lived low mass stars.  Because runaways ejected along the direction of Galactic 
rotation have higher ejection velocities and climb out of a shallower potential well 
than other runaways, these stars reach larger distances in the outer Galaxy. Although 
high velocity runaways appear at all $b_f$, the fastest unbound runaways are at low
galactic latitude, $b_f \lesssim$ 30\deg.

Comparisons of our results with observations of halo stars and simulations of HVSs
suggest clear differences between the three populations. Halo stars and HVSs are
uniformly distributed in $b$; runaways are concentrated in the disk. The radial 
density gradients of halo stars and HVSs are shallower than those of runaways. The
\vrad\ distributions of HVSs and runaways are clearly non-gaussian compared to 
observations of halo stars; HVSs tend to have larger \vrad\ than runaways. These 
differences suggest clear observational discriminants of the populations, which we
explore in the following sections.

\section{Application to the Hyper-Runaway HD~271791}

For a first application of our simulations, we consider the runaway B star HD~271791.
This B2--3 III star lies well below the Galactic plane ($z \approx -$10~kpc) and has 
a large heliocentric radial velocity of 442~\kms\ \citep{kilkenny88,kilkenny89}.
The observed proper motions suggest a large range in Galactic rest-frame velocity,
530--920~\kms, and an origin in the outer disk at $r \approx$ 12--16~kpc \citep{heber08}. 
Detailed abundance analyses yield a subsolar iron abundance and enhanced abundances 
of the $\alpha$-capture nuclei O, Ne, and S \citep{przybilla08a}. The subsolar iron
abundance supports an origin in the outer Galaxy; the high abundances of $\alpha$-nuclei
imply contamination of the atmosphere from a nearby supernova.

There are two proposals for the origin of HD~271791 as a hyper-runaway star.
\citet{przybilla08a} suggest that HD~271791 is the secondary of a massive binary 
disrupted by a supernova. However, the upper limit of the observed rest-frame velocity, 
$\sim$ 900 \kms, is hard to achieve in a close binary system. Thus \citet{gvara09a} 
prefers ejection during a three-body or four-body encounter in the dense core of a 
massive star cluster. 

The large range of allowed rest-frame velocities does not permit a unique interpretation
for this star. Here, we first consider the minimum rest-frame velocity of 530 \kms\ and 
note how larger velocities impact our conclusions. 

For a rest-frame velocity of 530 \kms, HD~271791 is a plausible runaway star 
produced by a supernova explosion in a massive binary system. With a distance
$r \approx$ 21 kpc from the \GC\ and a height $z \approx -$10 kpc below the Galactic 
plane \citep{heber08}, the star is marginally bound to the Galaxy using the 
\citet{kenyon08} potential model for the Milky Way. The radial velocity is similar 
to the maximum velocity achieved in our simulations (Figure~\ref{fig:rmv4-h}).  
The galactic latitude of $b \approx$ $-$25\deg\ is marginally consistent with the 
runaway interpretation.  Because the highest velocity runaways are ejected into 
the disk, only $\approx$ 5\%--10\% of unbound runaways have $b_f \gtrsim$ 
20\deg\ (Figure~\ref{fig:z-prob}).  Roughly 30\% of unbound runaways are 
ejected from the apparent origin of HD~271791 at $r \approx$ 12--16 kpc 
(Figure~\ref{fig:cum-prob}).  After the ejection, an 11 \msun\ star can reach 
$z \approx$ 10 kpc during its main sequence lifetime of 20--30 Myr 
\citep[Table 3, and][]{heber08}.  Thus, the observed spectral type is also
consistent with the runaway interpretation.

Larger rest-frame velocities weaken the case for the supernova ejection model.  In our 
simulations, the largest observed rest-frame radial velocity is roughly 100~\kms\ larger 
than the maximum ejection velocity from the binary. An observed rest-frame velocity of 
$\sim$ 900~\kms\ requires an ejection velocity of 800~\kms, roughly a factor of two 
larger than the maximum achieved in numerical simulations 
\citep[][see also Przybilla et al 2008]{portegies00}. Off-center supernova 
explosions might enhance the ejection velocity, but factor of two increases
for a massive star like HD~271791 seem unlikely \citep{gvara09a}.

Dynamical ejection mechanisms can explain the high rest-frame velocity of HD~271791.
\citet{gvara09a} outlines several mechanisms where the interactions among binary
or triple systems produce an ejected star with a velocity of $\lesssim$ 800 \kms.  To 
account for the enhanced $\alpha$-nuclei in the atmosphere, the dynamical encounter(s) 
in each mechanism must occur close in time to the supernova explosion of one of the 
companion stars to HD~271791. 

Low probabilities complicate all formation mechanisms for HD~271791 \citep{brown09a}. 
In the \citet{przybilla08a} model, a supernova explosion in a 80 \msun\ star produces
the high ejection velocity of the secondary.  For a Galactic star formation rate of 
4 \msun\ yr$^{-1}$ \citep[e.g.,][]{diehl06}, we expect $\approx 5 \times 10^4$ stars 
with \mstar\ $\gtrsim$ 80 \msun\ during the 25 Myr main sequence lifetime of HD~271791.  
Roughly one-third of all O star binaries are twins; for a \citet{salpeter55} initial mass
function, 5\% of the rest have 10--12~\msun\ companions \citep{kobul07}.  Thus, $\sim$ 
1000 massive binaries with
80 \msun\ primary stars have 10--12~\msun\ secondaries. If all supernova ejections in 
these binaries produce a runaway, we expect 1000 $\times$ 0.08\% $\approx$ 1 unbound runaway 
similar to HD~271791 every 25 Myr. Extrapolating the results of our simulations to 
11 \msun\ runaways, the joint probability of observing this runaway at $b \approx$ 
$-$25\deg\ and at the end of its main sequence lifetime at $r \approx$ 20 kpc is $\sim$ 
0.1 (see Table \ref{tab:med-zr}). Thus, HD~271791 is an unlikely runaway star.

\citet{brown09a} show that the likelihood of observing a hyper-runaway from a dynamical 
ejection is also very small. Dynamical ejections require interactions between two binary
systems composed of massive stars. Thus, the probability of a dynamical ejection is the 
joint probability of interacting masive pairs of binaries within the stars' main sequence 
lifetimes.  In an ensemble of dense clusters capable of producing high velocity ejections, 
the probability of a 3--4~\msun\ hyper-runaway is $\sim 10^{-5}$.  For a Salpeter (1955)
initial mass function, 11~\msun\ stars are $\sim$ 4--5 times less likely than 3--4~\msun\ stars. 
Thus, dynamical ejections are much less likely than supernova ejections.

An improved proper motion for HD~271791 would place better constraints on formation
mechanisms. Current data have large uncertainties, leading to a large range in rest-frame
velocity \citep{heber08}.  Observations with GAIA, scheduled for a 2011 launch, would 
yield a very accurate rest-frame velocity. 

\section{Application to Hypervelocity Stars}

	We now explore the observational consequences of runaway star distributions
for HVS surveys.  First, we compare our runaway simulations to numerical simulations 
of the apparent magnitude and heliocentric radial velocity distributions of HVSs.  
We then compare our runaway simulations with the HVS observations of Brown et al.
Finally, we estimate the possible contribution of runaways to the observed sample of HVSs.

\subsection{Predicting Observables from Simulations}

	We begin by ``observing'' our numerical simulations for 3 \Msolar\ runaways 
and HVSs in a heliocentric reference frame.  
Known HVSs come mostly from the radial velocity survey of \citet{brown06, brown06b,
brown07a, brown07b, brown09a} who target objects with the colors of 3--4 \Msolar\ stars.
Three of the HVSs are confirmed $\simeq$3 \Msolar\ main sequence stars 
\citep{ fuentes06, przybilla08b, lopezmorales08}. \citet{girardi02, girardi04}
stellar evolutionary tracks show that a solar metallicity, 3 \Msolar\ star spends
350 Myr on the main sequence with an average luminosity of $M_g=0.0$.

	Thus we calculate heliocentric distances and apparent magnitudes for the
simulated runaways assuming $M_g=0.0$ appropriate for a 3 \Msolar\ star.  We assume
the Sun is located at $r=8$ kpc.  We shift the origin of the coordinate system to
the Sun, and derive heliocentric radial velocities by taking $\vec{v} \cdot \hat{r}$.

\begin{figure}
% \centerline{\includegraphics[width=6.0in]{hvs33.ps}}
 \plotone{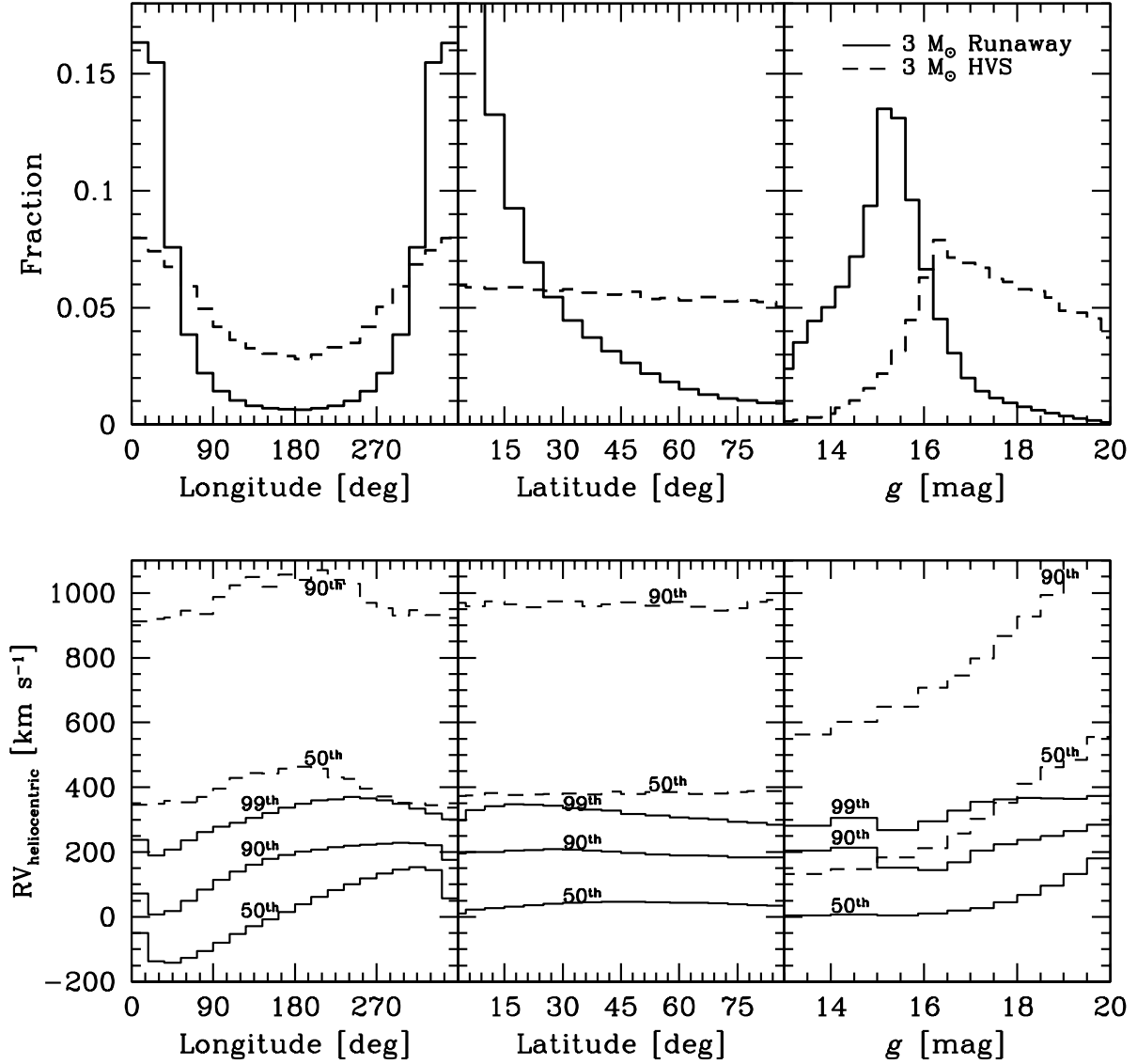}
\caption{ \label{fig:hvs3}
	Predicted distributions of runaways and HVSs in a heliocentric reference
frame. The top panels plot number fractions of simulated runaways ({\it solid lines}) 
and HVSs ({\it dashed lines}) as a function of Galactic longitude, latitude, and 
apparent magnitude.  
Bottom panels plot the 50$^{\rm th}$, 90$^{\rm th}$, and 99$^{\rm th}$
percentile heliocentric radial velocities of simulated runaways and HVSs.  
The spatial and velocity distributions of 3 \Msolar\ runaways and HVSs are 
very different: runaways are apparently bright and concentrated to the Galactic
plane; HVSs are apparently faint and common in the halo. 
}
  \end{figure}

	Figure \ref{fig:hvs3} plots a heliocentric view of the simulations as a
function of Galactic longitude, latitude, and apparent magnitude. The top panels of
Figure \ref{fig:hvs3} plot the number distribution of runaways and HVSs. The bottom 
panels of Figure \ref{fig:hvs3} plot the 50$^{\rm th}$, 90$^{\rm th}$, and 99$^{\rm 
th}$ percentile heliocentric radial velocity of runaways and HVSs.

	The spatial concentration of runaways in the disk, discussed in \S2 above, is 
evident in Figure \ref{fig:hvs3}.  Our models predict a greater fraction of runaways 
in the Galactic center hemisphere $|l| < \pm 90^{\circ}$ and at low Galactic latitudes
$b<30^{\circ}$.  The fraction of HVSs, on the other hand, is larger than the fraction
of runaways in the direction of the Galactic anti-center $l=180^{\circ}$ and the Galactic 
pole $|b|=90^{\circ}$.  Compared to HVSs, runaways are also apparently bright: 80\% of 
3~\Msolar\ runaways are {\it brighter} than $g=16$, whereas 85\% of 3~\Msolar\ HVSs are 
{\it fainter} than $g=16$.

	Runaways are ejected from a rotating disk. This rotation is apparent in the 
distribution of heliocentric radial velocities:  the median (50$^{\rm th}$ percentile) 
runaway velocity is negative in the direction of Galactic rotation $0^{\circ}< l <180^{\circ}$ 
and positive in the direction opposite Galactic rotation $180^{\circ}< l <360^{\circ}$.  
Similarly, the latitude dependence of runaway velocities reflects the boost from Galactic 
rotation at low latitudes.  HVSs, on the other hand, are ejected on purely radial trajectories 
and show no rotation.  Median HVS velocities exceed the 99$^{\rm th}$ percentile runaway 
velocity in every direction on the sky.  Fainter stars are faster because only the fastest 
runaways and HVSs survive to reach the largest distances.

\subsection{Comparing Simulations and Observations}

	We now compare our runaway simulation of 3 \Msolar\ stars to observations of
HVSs.  Observed HVSs are significant velocity outliers with minimum radial
velocities in the Galactic rest frame $>$~+400 \kms.  The well-defined survey of
\citet{brown06, brown06b, brown07a, brown07b, brown09a} samples stars with
magnitudes $15<g<20.5$ and latitudes $30^{\circ}\lesssim b <90^{\circ}$. 

	Over the range of magnitude and latitude sampled by the Brown et al survey,
only 0.001\% of the simulated 3 \Msolar\ runaways have velocities $>$~+400 \kms. 
These runaways differ in three ways from the observed HVSs.  (1) Simulated runaways 
with radial velocities exceeding $400$ \kms\ are located at low latitude with median 
$b=34^{\circ}$;  observed HVSs are distributed uniformly across Galactic latitude with 
median $b=51^{\circ}$ \citep{brown09b}. (2) The $>400$ \kms\ simulated runaways are 
bright with median $g=16$;  observed HVSs are faint with median $g=19$ \citep{brown09a}.
(3) The fastest simulated runaway with $b>30^{\circ}$ has velocity +450 \kms;
observed HVSs have velocities up to +700 \kms.  We conclude that runaways cannot
significantly contaminate the HVS samples because the observed distribution of HVSs
differs so markedly from that expected for runaways (Figure \ref{fig:hvs3}).

	We now consider how many runaways with radial velocities exceeding +400 \kms\ might 
be included in the Brown et al HVS survey.  To explore this point, we normalize the number 
of runaways in our simulation to the total number of 3 \Msolar\ stars formed in the last 
350 Myr.  Assuming the star formation rate in the Galactic disk is 4 M$_{\sun}$ yr$^{-1}$
\citep{diehl06}, $1.4\times10^9$ \Msolar\ of stars have formed in the disk in the
past 350 Myr.  A Salpeter initial mass function, integrated from 0.1 to 100 \Msolar\
and normalized to $1.4\times10^9$ \Msolar, predicts $1.3\times10^7$ 3 - 4 \Msolar\
stars.  Assuming that $\sim$1\% of all stars are ejected as runaways (see below), we
predict $\sim$1 3 \Msolar\ runaway with +400 \kms\ radial velocity in the Brown
et~al.\ HVS survey.  This prediction is comparable to the analytic estimate in
\citet{brown09a}, and suggests that perhaps one of the observed HVSs may be a
runaway.  

\section{Application to Halo Radial Velocity Surveys}

	Large modern photometric and radial velocity surveys open the
possibility of global constraints on the number and distribution of
runaways. In this section, we consider several examples. We first compare
the predictions of our runaway star simulations with the stellar population 
of the Milky Way derived from color selected star counts in the SDSS. This 
comparison indicates where future searches for runaways would be most productive. 
Combined with a 2MASS-selected spectroscopic survey \citep{brown08}, the SDSS counts 
provide an upper limit on the fraction of disk A main sequence stars ejected as runaways.

	Our numerical simulations demonstrate that runaways reach the inner halo of the
Milky Way. Their velocity dispersion is comparable to the halo velocity dispersion; their 
rotation is comparable with the rotation of the thick disk. Thus runaways are difficult to 
identify by their kinematics, but their high metallicities should distinguish them from 
typical halo stars.

\subsection{Comparing Runaways to Star Counts}

	We begin by comparing our runaway simulations to the observed
stellar population of the Milky Way as revealed by star counts in the
Sloan Digital Sky Survey (SDSS) Data Release 6 (DR6) \citep{adelman08}.
For this comparison, we select simulations for 1.5 \Msolar, 2 \Msolar, 
and 3~\Msolar\ stars.  As before, we ``observe'' the simulations from a
heliocentric reference frame. We calculate apparent magnitudes assuming the 
1.5 \Msolar, 2 \Msolar, and 3~\Msolar\ runaways have main sequence luminosities 
of $M_g=+2.9$, +1.5, and 0.0, respectively \citep{girardi02, girardi04}.  

	To compare observed number counts of stars with predictions from our
numerical simulations, we consider only runaways that fall in the region of 
sky imaged by the SDSS DR6.  For the stars in the SDSS, we count those stars 
with colors $0.15<(g-r)_0<0.20$, $-0.15<(g-r)_0<-0.05$, and $-0.35<(g-r)_0<-0.25$ 
appropriate for 1.5 \Msolar, 2 \Msolar, and 3 \Msolar\ stars, respectively, 
according to the \citet{girardi02, girardi04} stellar evolutionary tracks for 
solar metallicity stars.

\begin{figure}
% \centerline{\includegraphics[width=4.0in]{runaway2.ps}}
 \plotone{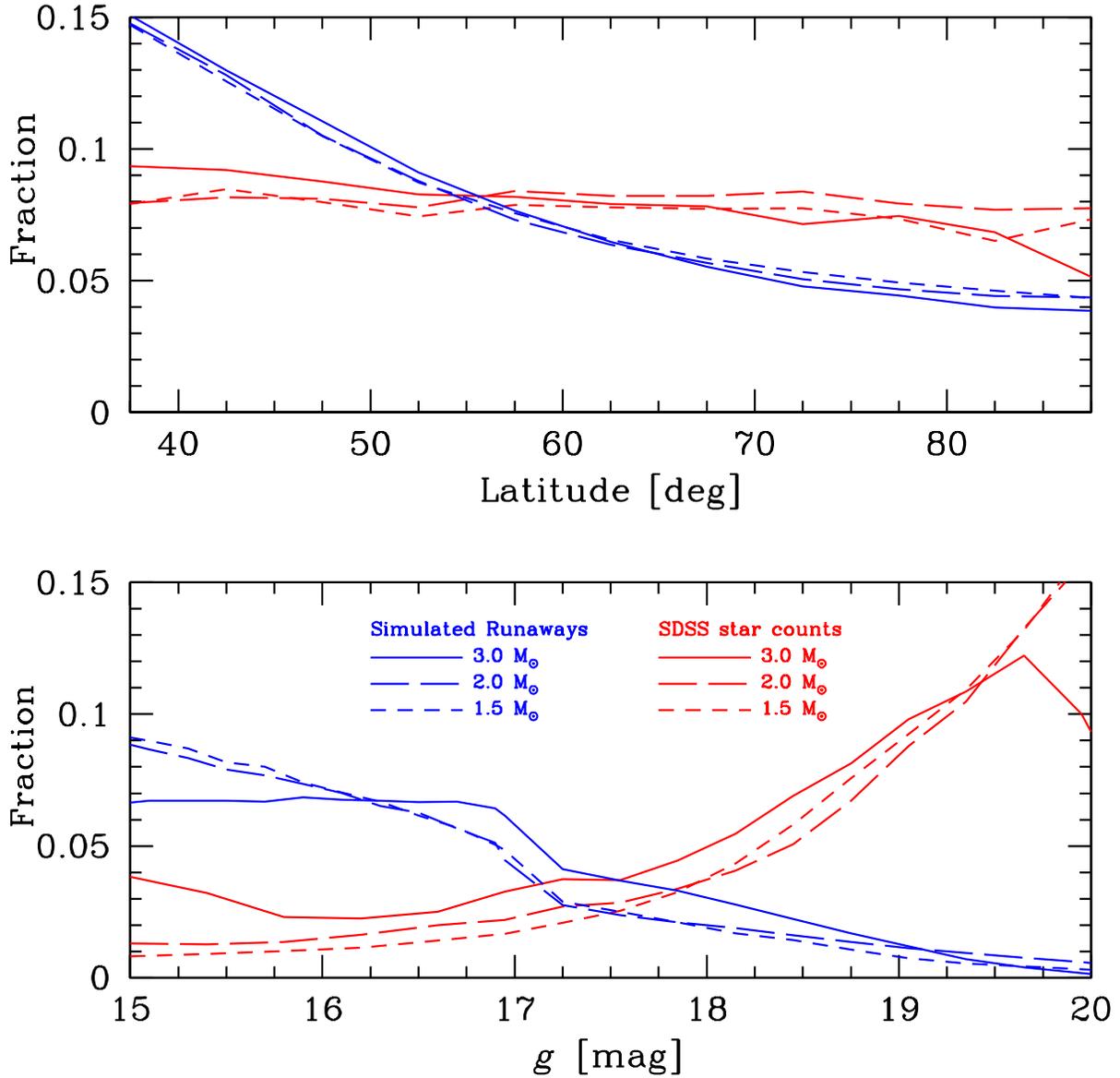}
\caption{ \label{fig:runaway}
	Comparison of number fractions of simulated 1.5, 2 and 3 \Msolar\ runaways 
in the SDSS survey region ({\it blue lines}) with SDSS star counts ({\it red lines}).
Relative to the Galactic stellar population, runaways have the highest contrast
at low latitudes and at bright magnitudes.}
  \end{figure}

	Figure \ref{fig:runaway} plots the resulting number fraction of runaways and
SDSS stars as a function of Galactic latitude and apparent magnitude.  Over the
region surveyed by SDSS, a larger fraction of runaways are found at low latitudes
compared to star counts.  This latitude dependence reflects the flattened distribution 
of runaways compared to the population of halo stars that dominate the star counts.  
Runaways contribute a negligible amount to the observed stellar (halo) population at
faint magnitudes, $g\gtrsim17$.

\subsection{Upper Limit on the Runaway Fraction}

	The runaway fraction of O- and B-type stars has long been known to be $\sim$40\% 
and $\sim$5\%, respectively \citep{blaauw61, gies86}. However, there are few comparable 
constraints on the fraction of A-type runaways.  \citet{stetson81} used tangential velocity 
cuts of bright stars with rough spectral types to estimate that at most $\sim$0.3\% of solar 
neighborhood A-type stars are runaways.  Here, we combine constraints from our simulations
and existing spectroscopic surveys to place an independnent upper limit on the fraction of 
2~\Msolar\ stars ejected as runaways. 

To make this estimate, we use the \citet{brown08} spectroscopic survey of A-type stars 
in the Two Micron All Sky Survey \citep{skrutskie06}.  The \citet{brown08} survey is 
complete over 4300 deg$^{2}$ to a magnitude limit of $J_0=15.5$, equivalent to $g\simeq15.5$
for a zero-color A star. \citet{brown08} find that 40\% of A-type stars at 15$^{\rm
th}$ mag are main sequence stars (the other 60\% are evolved horizontal branch
stars). A third of the main sequence stars are consistent with having solar
metallicity, thus 13\% of 15$^{\rm th}$ mag A stars are possible runaways.  In our
2 \Msolar\ simulation, however, the fraction of runaways is six times larger than
than the observed fraction in star counts at $g=15$ (Figure \ref{fig:runaway}).  Thus, 
if the main sequence A stars observed by \citet{brown08} are all runaways, the absolute
fraction of A stars ejected as runaways is no larger than $0.13/6 \simeq 2\%$, in
agreement with the upper limit estimated by \citet{stetson81}.

\subsection{Runaways in the Inner Halo}

	We now broaden our discussion and look at the distribution of runaways 
in the context of Galactic structure.

	Spatially, 80\% of runaways in our simulations are located at $|b|<15^{\circ}$
(Figure \ref{fig:runaway}), a population an observer might call ``thick disk.'' The
Galactic thick disk has an observed scale height of 0.75 - 1.0 kpc
\citep[i.e.][]{siegel02, girard06, juric08}, and it dominates the number density of
stars in the region $1<|z|<5$ kpc.  We note that claims of unusually large thick disk
scale heights are possibly confused with the inner stellar halo \citep{kinman09}. By
comparison, the number density of runaways in our simulations, selected with $7<r_{\rm
cyl}<9$ kpc and $1<|z|<5$ kpc, is well fit by an exponential distribution with a
vertical scale height $h_z=0.54\pm0.01$ kpc.

	Kinematically, runaways are a hot, rotating population analogous to the
thick disk and inner stellar halo.  In the region away from the plane
$|b|>15^{\circ}$, simulated runaways have a mean 135 \kms\ component of velocity in
the direction of Galactic rotation, comparable to that observed for the thick disk
\citep{chiba00}.  Runaways also have a 130 \kms\ heliocentric radial velocity
dispersion, essentially identical to the velocity dispersion of inner halo stars
with the same apparent magnitude.  The Milky Way inner halo, as described by
\citet{carollo07} and \citet{morrison09}, has a small $\sim$25 \kms\ prograde
rotation and a 1D velocity dispersion of $\sim$120 \kms .  Thus runaways have
similar kinematics to the Galactic thick disk and inner halo and are difficult to
identify by radial velocity alone.

	The mean metallicity of the inner halo is [Fe/H]$=-1.6$ \citep{carollo07}, but
\citet{ivezic08} report solar metallicity stars up to 5 kpc above the plane.  Perhaps
runaways can contribute to the high metallicity population and to the small prograde
rotation observed in the inner stellar halo.

	The inner halo is much too metal poor to be composed entirely of runaways, yet
metal rich, short-lived runaways should be present in the halo of the Milky Way.  The
deaths of massive runaways must necessarily result in metal enrichment and energy
input into the halo.  This conclusion applies for all star-forming galaxies, at all
redshifts.  In particular, runaways should be more abundant early in the evolution of
a galaxy when the star formation rate is larger. Thus, the distribution of runaways
may have important implications for feedback processes.

\section{Conclusions}

	Runaway stars are an interesting class of objects because they
connect star formation in the disk with the halo of the Milky Way. We
explore these connections by using the \citet{portegies00} distribution
of binary-supernova ejections to inject stars into the Galactic potential.
We track the progress of these stars from the Galactic disk to the Galactic
halo and derive simulated catalogs of runaways.

	We show that the velocity and spatial distributions of runaways
depend on the Galactic potential and stellar lifetime. All runaways have
a flattened spatial distribution, with higher velocity runaways at lower
Galactic latitudes. Massive runaways do not live long enough to reach the
outer halo. Thus, massive runaways are more concentrated towards the
Galactic center and in the plane of the disk than low mass runaways.

	Kinematically, runaways are a hot, rotating population of stars with
dynamical properties between the thick disk and the halo. In the solar
neighborhood, runaways with masses of 1.5--3~\msun\ have scale heights and
rotation velocities comparable to the thick disk, and velocity dispersions
comparable to the inner stellar halo.  Although they do not have a unique
signature in radial velocity surveys, runaways are overwhelmingly located at
low Galactic latitudes and at bright apparent magnitudes. Our results suggest
an upper limit of 2\% of A-type stars ejected as runaways.

	The kinematics of the unbound runaway HD 271791 provides an interesting 
comparison with the simulations. For runaways that reach distances of 60--100 kpc 
from the Galactic Center, the simulations predict a peak in the initial distance 
of bound (unbound) stars at 7--12 kpc (10--15 kpc). \citet{heber08} conclude that
HD 271791 originated from 12~kpc $\lesssim r \lesssim16$ kpc \citep{heber08}. 
Clear tests of the simulations require larger samples of runaways with high
quality proper motions and good estimates for their starting locations in the disk.

	Radial velocity surveys for high velocity outliers are unlikely to
confuse runaway stars with HVSs. Among ejected stars with velocities
exceeding 400 \kms, runaways (1) are brighter ($g \lesssim$ 18),
(2) have smaller velocities ($v \lesssim$ 450 \kms), and (3) are more
concentrated to lower galactic latitudes ($b \lesssim 35^{\circ}$) than HVS
with typical $g \gtrsim$ 18, $v \approx$ 400--700 \kms, and random $b$ in
the range $30^{\circ}< b <90^{\circ}$. We estimate that at most $\sim 1$
runaway contaminates the Brown et~al.\ sample of HVSs.

	Future progress on the theory of runaways requires predictions for
the velocity distribution from the dynamical ejection mechanism. Although
this process can yield higher ejection velocities than the
binary-supernova ejection mechanism, uncertainties in the stellar merger
rate during close encounters complicates calculations of a realistic
maximum ejection velocity and a distribution of ejection velocities.

	For both mechanisms, better estimates of predicted rates are
needed to constrain predictions for the frequency and kinematics of
runaways. Extending the \citet{portegies00} simulations to lower mass
stars would improve our estimate for the rate of runaways from the
binary-supernova ejection mechanism among A-type stars.  Simulations of
ensembles of dense star clusters would yield ejections rates for the
dynamical ejection mechanism.

	Measurements of proper motions and metallicities of inner halo
stars can also improve our understanding of runaways. Identifying the
fraction of halo stars with roughly solar metallicity can yield a better
estimate of the frequency of runaways. Proper motions and radial
velocities place clear constraints on the origin of runaways for
comparison with theoretical simulations.

\acknowledgements

	We acknowledge Elliott Barcikowsky's contribution to the early
stages of this work.  Comments from an anonymous referee greatly improved
our presentation. This research makes use of NASA's Astrophysics Data
System Bibliographic Services and data products from the Sloan Digital Sky 
Survey, which was funded by the Alfred P.\ Sloan Foundation and Participating 
Institutions.  This work was supported in part by the Smithsonian Institution.

\clearpage

% \bibliographystyle{/home/wbrown/lib/apj}
% \bibliography{/home/wbrown/text/RefHS}

\begin{deluxetable}{ccccccc}
\tablecolumns{7}
\tablewidth{0pc}
\tabletypesize{\footnotesize}
\tablenum{1}
\tablecaption{Median Radial Velocities of Runaway Stars \tablenotemark{a}}
\tablehead{
  \colhead{ \ } &
  \multicolumn{6}{c}{Galactocentric distance $r$}
\\
  \colhead{$b_f$ (deg)} &
  \colhead{$<$ 20 kpc} &
  \colhead{20--40 kpc} &
  \colhead{40--60 kpc} &
  \colhead{60--80 kpc} &
  \colhead{80--100 kpc} &
  \colhead{100--120 kpc}
}
\startdata
\cutinhead{\mstar\ = 1.5 \msun, $t_{ms}$ = 2.9 Gyr, \rinit\ = 10 kpc, \vej = 400 \kms}
0--30 & ~20 & ~15 & ~26 & ~25 & ~31 & ~44 \\
30--50& ~69 & $-$35 & $-$46 & $-$37 & ~10 & ~23 \\
50--90& $-$13 & ~94 & ~92 & ~51 & ~34 & ~23 \\
\cutinhead{\mstar\ = 2 \msun, $t_{ms}$ = 1.2 Gyr, \rinit\ = 10 kpc, \vej = 400 \kms}
0--30 & ~45 & 54 & ~83 & ~78 & 150 & 170 \\
30--50& 110 & 44 & $-$33 & $-$27 & ~40 & ~58 \\
50--90& ~10 & 93 & 111 & ~62 & ~35 & ~24 \\
\cutinhead{\mstar\ = 3 \msun, $t_{ms}$ = 350 Myr, \rinit\ = 3 kpc, \vej = 400 \kms}
0--30 & ~86 & 122 & 142 & 133 & 150 & 158 \\
30--50& $-$36 & ~22 & ~40 & ~37 & \nodata & \nodata \\
50--90& 100 & ~33 & \nodata & \nodata & \nodata & \nodata \\
\cutinhead{\mstar\ = 3 \msun, $t_{ms}$ = 350 Myr, \rinit\ = 10 kpc, \vej = 400 \kms}
0--30 & 112 & 190 & 240 & 275 & 295 & 304 \\
30--50& 222 & 266 & 230 & 210 & 205 & 230 \\
50--90& 166 & 190 & 152 & 115 & 130 & \nodata \\
\cutinhead{\mstar\ = 3 \msun, $t_{ms}$ = 350 Myr, \rinit\ = 30 kpc, \vej = 400 \kms}
0--30 &$-$36& 112 & 314 & 339 & 363 & 384 \\
30--50&$-$26& 165 & 281 & 321 & 330 & 333 \\
50--90& ~60 & 145 & 226 & 254 & 259 & 258 \\
\cutinhead{\mstar\ = 3 \msun, $t_{ms}$ = 350 Myr, \rinit\ = 10 kpc, \vej = 350 \kms}
0--30 & ~90 & 155 & 207 & 228 & 232 & 243 \\
30--50& 197 & 191 & 104 & 125 & 155 & 176 \\
50--90& ~98 & 127 & ~74 & ~42 & \nodata & \nodata \\
\cutinhead{\mstar\ = 4 \msun, $t_{ms}$ = 160 Myr, \rinit\ = 10 kpc, \vej = 400 \kms}
0--30 & 223 & 320 & 361 & 382 & 404 & \nodata \\
30--50& 249 & 308 & 295 & 319 & \nodata & \nodata \\
50--90& 201 & 232 & 221 & \nodata & \nodata & \nodata \\
\cutinhead{\mstar\ = 5 \msun, $t_{ms}$ = 95 Myr, \rinit\ = 10 kpc, \vej = 400 \kms}
0--30 & 254 & 388 & 418 & \nodata & \nodata & \nodata \\
30--50& 260 & 335 & 359 & \nodata & \nodata & \nodata \\
50--90& 217 & 270 & 301 & \nodata & \nodata & \nodata \\
\cutinhead{\mstar\ = 6 \msun, $t_{ms}$ = 65 Myr, \rinit\ = 10 kpc, \vej = 400 \kms}
0--30 & 271 & 418 & \nodata & \nodata & \nodata & \nodata \\
30--50& 268 & 351 & \nodata & \nodata & \nodata & \nodata \\
50--90& 239 & 292 & \nodata & \nodata & \nodata & \nodata \\
\enddata
\tablenotetext{a}{Results for fiducial simulations of $10^6$ stars 
with one initial starting radius, stellar mass, and ejection velocity 
for all stars.  Each entry lists the median \vrad, measured in the
Galactocentric reference frame, as a function of $r$ and $b_f$
for stars with the listed initial conditions.  For entries without 
data, simulations with $10^6$ trials do not produce stars with these 
combinations of $r$ and $b_f$. }
\label{tab:fidmed}
\end{deluxetable}
\clearpage

\begin{deluxetable}{ccccccc}
\tablecolumns{7}
\tablewidth{0pc}
\tabletypesize{\footnotesize}
\tablenum{2}
\tablecaption{Radial Velocity Dispersions of Runaway Stars \tablenotemark{a}}
\tablehead{
  \colhead{ \ } &
  \multicolumn{6}{c}{Galactocentric distance $r$}
\\
  \colhead{$b_f$ (deg)} &
  \colhead{$<$ 20 kpc} &
  \colhead{20--40 kpc} &
  \colhead{40--60 kpc} &
  \colhead{60--80 kpc} &
  \colhead{80--100 kpc} &
  \colhead{100--120 kpc}
}
\startdata
\cutinhead{\mstar\ = 1.5 \msun, \rinit\ = 10 kpc, \vej = 400 \kms}
0--30 & 193 & 204 & 177 & 146 & 140 & 144 \\
30--50& 200 & 212 & 190 & 168 & 154 & 135 \\
50--90& 184 & 158 & 121 & ~85 & ~70 & ~46 \\
\cutinhead{\mstar\ = 2 \msun, \rinit\ = 10 kpc, \vej = 400 \kms}
0--30 & 192 & 199 & 172 & 172 & 174 & 154 \\
30--50& 190 & 220 & 200 & 162 & 134 & 110 \\
50--90& 187 & 172 & 133 & ~90 & ~69 & ~46 \\
\cutinhead{\mstar\ = 3 \msun, \rinit\ = 3 kpc, \vej = 400 \kms}
0--30 & 179 & 171 & 108 & ~64 & ~30 & ~~4 \\
30--50& 206 & 134 & ~72 & ~27 & \nodata & \nodata \\
50--90& 140 & ~78 & \nodata & \nodata & \nodata & \nodata \\
\cutinhead{\mstar\ = 3 \msun, \rinit\ = 10 kpc, \vej = 400 \kms}
0--30 & 180 & 201 & 157 & 117 & ~83 & ~54 \\
30--50& 155 & 165 & 123 & ~84 & ~55 & ~33 \\
50--90& 160 & 136 & ~87 & ~48 & ~23 & \nodata \\
\cutinhead{\mstar\ = 3 \msun, \rinit\ = 30 kpc, \vej = 400 \kms}
0--30 & 257 & 180 & 137 & 128 & 109 & ~89 \\
30--50& 179 & 120 & 104 & 100 & ~82 & ~61 \\
50--90& 101 & ~87 & ~94 & ~77 & ~54 & ~34 \\
\cutinhead{\mstar\ = 3 \msun, \rinit\ = 10 kpc, \vej = 350 \kms}
0--30 & 165 & 183 & 134 & ~92 & ~58 & ~34 \\
30--50& 143 & 148 & 101 & ~62 & ~32 & ~~8 \\
50--90& 138 & 112 & ~61 & ~21 & \nodata & \nodata \\ 
\cutinhead{\mstar\ = 4 \msun, \rinit\ = 10 kpc, \vej = 400 \kms}
0--30 & 148 & 141 & ~85 & ~41 & ~~8 & \nodata \\
30--50& 124 & 108 & ~58 & ~27 & \nodata & \nodata \\
50--90& 111 & ~83 & ~36 & \nodata & \nodata & \nodata \\
\cutinhead{\mstar\ = 5 \msun, \rinit\ = 10 kpc, \vej = 400 \kms}
0--30 & 130 & 103 & ~46 & \nodata & \nodata & \nodata \\
30--50& 100 & ~74 & ~31 & \nodata & \nodata & \nodata \\
50--90& ~82 & ~52 & ~10 & \nodata & \nodata & \nodata \\
\cutinhead{\mstar\ = 6 \msun, \rinit\ = 10 kpc, \vej = 400 \kms}
0--30 & 127 & ~85 & \nodata & \nodata & \nodata & \nodata \\
30--50& ~88 & ~61 & \nodata & \nodata & \nodata & \nodata \\
50--90& ~68 & ~35 & \nodata & \nodata & \nodata & \nodata \\
\enddata
\tablenotetext{a}{Results for fiducial simulations with one initial
starting radius and ejection velocity for each stellar mass.  Each 
entry lists the 1D radial velocity dispersion as a function of $r$ 
and $b_f$ for stars with the listed initial conditions.  For entries 
without data, the simulations do not produce stars with these 
combinations of $r$ and $b_f$. }
\label{tab:fiddisp}
\end{deluxetable}
\clearpage

\begin{deluxetable}{ccccccc}
\tablecolumns{7}
\tablewidth{0pc}
\tabletypesize{\footnotesize}
\tablenum{3}
\tablecaption{Median Radial Velocities of Runaway Stars Ejected from the Galactic Disk\tablenotemark{a}}
\tablehead{
  \colhead{ \ } &
  \multicolumn{6}{c}{Galactocentric distance $r$}
\\
  \colhead{Stellar Mass} &
  \colhead{$<$ 20 kpc} &
  \colhead{20--40 kpc} &
  \colhead{40--60 kpc} &
  \colhead{60--80 kpc} &
  \colhead{80--100 kpc} &
  \colhead{100--120 kpc}
}
\startdata
\cutinhead{$b < 30^\circ$}
1.5 M$_\odot$& ~~1 & ~~3 & ~~7 & ~~9 & ~15 & ~16 \\
2   M$_\odot$& ~~2 & ~~8 & ~17 & ~29 & ~28 & ~42 \\
3   M$_\odot$& ~~8 & ~23 & ~84 & 136 & 188 & 239 \\
4   M$_\odot$& ~20 & 114 & 239 & 332 & 411 & \nodata \\
5   M$_\odot$& ~53 & 202 & 361 & \nodata & \nodata & \nodata \\
6   M$_\odot$& ~72 & 246 & 355 & \nodata & \nodata & \nodata \\
\cutinhead{$ 30^\circ \le b \le 50^\circ $}
1.5 M$_\odot$& ~~4 & ~10 & ~14 & ~15 & ~16 & ~23 \\
2   M$_\odot$& ~13 & ~17 & ~16 & ~22 & ~27 & ~38 \\
3   M$_\odot$& ~50 &  ~46 & ~73 & 121 & 172 & 224 \\
4   M$_\odot$& ~70 &  115 & 217 & 309 & \nodata & \nodata \\
5   M$_\odot$& 103 &  199 & 334 & \nodata & \nodata & \nodata \\
6   M$_\odot$& 144 &  260 & \nodata & \nodata & \nodata & \nodata \\
\cutinhead{$b > 50^\circ$}
1.5 M$_\odot$& ~22 & ~48 & ~54 & ~52 & ~68 & ~49 \\
2   M$_\odot$& ~37 & ~64 & ~70 & ~71 & ~65 & ~48 \\
3   M$_\odot$& ~72 & ~88 & ~88 & 108 & 152 & 208 \\
4   M$_\odot$& ~97 & 127 & 194 & 285 & \nodata & \nodata \\
5   M$_\odot$& 128 & 187 & 287 & \nodata & \nodata & \nodata \\
6   M$_\odot$& 169 & 245 & \nodata & \nodata & \nodata & \nodata \\
\enddata
\tablenotetext{a}{Results for $10^7$ stars ejected with velocity \vinit\ from 
$r$ = \rinit\ in the Galactic Disk. The ejection velocity and position are chosen from 
probability distributions (Eq. \ref{eq:pdfr}--\ref{eq:pdfv}) described in the main text.
Each column lists the median \vrad\ for 1.5--6 \msun\ stars with $r$ in the listed range.}
\label{tab:fullmed}
\end{deluxetable}
\clearpage

\begin{deluxetable}{ccccccc}
\tablecolumns{7}
\tablewidth{0pc}
\tabletypesize{\footnotesize}
\tablenum{4}
\tablecaption{Radial Velocity Dispersions of Runaway Stars Ejected from the Galactic Disk\tablenotemark{a}}
\tablehead{
  \colhead{ \ } &
  \multicolumn{6}{c}{Galactocentric distance $r$}
\\
  \colhead{Stellar Mass} &
  \colhead{$<$ 20 kpc} &
  \colhead{20--40 kpc} &
  \colhead{40--60 kpc} &
  \colhead{60--80 kpc} &
  \colhead{80--100 kpc} &
  \colhead{100--120 kpc}
}
\startdata
\cutinhead{$b < 30^\circ$}
1.5 M$_\odot$& 102 &  127 & 129 & 124 & 116 & 115 \\
2   M$_\odot$& ~99 &  125 & 126 & 124 & 122 & 112 \\
3   M$_\odot$& ~94 &  125 & 109 & ~86 & ~66 & ~50 \\
4   M$_\odot$& ~93 &  102 & ~71 & ~44 & ~32 & \nodata \\
5   M$_\odot$& ~94 &  102 & ~53 & \nodata & \nodata & \nodata \\
6   M$_\odot$& 104 &  125 & ~62 & \nodata & \nodata & \nodata \\
\cutinhead{$ 30^\circ \le b \le 50^\circ $}
1.5 M$_\odot$& 121 &  127 & 117 & 116 & 117 & 111 \\
2   M$_\odot$& 120 &  127 & 125 & 118 & 106 & ~97 \\
3   M$_\odot$& 118 &  118 & ~96 & ~73 & ~55 & ~40 \\
4   M$_\odot$& 102 &  ~90 & ~58 & ~34 & \nodata & \nodata \\
5   M$_\odot$& ~88 &  ~74 & ~41 & \nodata & \nodata & \nodata \\
6   M$_\odot$& ~83 &  ~67 & \nodata & \nodata & \nodata & \nodata \\
\cutinhead{$b > 50^\circ$}
1.5 M$_\odot$& 121 &  114 & ~96 & ~84 & ~87 & ~82 \\
2   M$_\odot$& 121 &  110 & ~99 & ~91 & ~88 & ~80 \\
3   M$_\odot$& 117 &  104 & ~79 & ~56 & ~40 & ~25 \\
4   M$_\odot$& 103 &  ~76 & ~42 & ~15 & \nodata & \nodata \\
5   M$_\odot$& ~82 &  ~54 & ~16 & \nodata & \nodata & \nodata \\
6   M$_\odot$& ~67 &  ~36 & \nodata & \nodata & \nodata & \nodata \\
\enddata
\tablenotetext{a}{As in Table \ref{tab:fullmed} for the radial velocity dispersion.}
\label{tab:fulldisp}
\end{deluxetable}
\clearpage

\begin{deluxetable}{ccccccc}
\tablecolumns{7}
\tablewidth{0pc}
\tabletypesize{\footnotesize}
\tablenum{5}
\tablecaption{Vertical Velocity Dispersions of Runaway Stars Ejected from the Galactic Disk\tablenotemark{a}}
\tablehead{
  \colhead{ \ } &
  \multicolumn{6}{c}{Galactocentric distance $r$}
\\
  \colhead{Stellar Mass} &
  \colhead{$<$ 20 kpc} &
  \colhead{20--40 kpc} &
  \colhead{40--60 kpc} &
  \colhead{60--80 kpc} &
  \colhead{80--100 kpc} &
  \colhead{100--120 kpc}
}
\startdata
1.5 M$_\odot$& 65 &  60 & 49 & 42 & 36 & 35 \\
2   M$_\odot$& 64 &  61 & 51 & 47 & 42 & 37 \\
3   M$_\odot$& 67 &  67 & 50 & 41 & 41 & 42 \\
4   M$_\odot$& 70 &  62 & 57 & 58 & 52 & \nodata \\
5   M$_\odot$& 69 &  69 & 68 & \nodata & \nodata & \nodata \\
6   M$_\odot$& 71 &  78 & 74 & \nodata & \nodata & \nodata \\
\enddata
\tablenotetext{a}{As in Table \ref{tab:fullmed} for the velocity dispersion 
perpendicular to the disk.}
\label{tab:fullvdisp}
\end{deluxetable}
\clearpage

\begin{deluxetable}{ccccccc}
\tablecolumns{7}
\tablewidth{0pc}
\tabletypesize{\footnotesize}
\tablenum{6}
\tablecaption{Vertical Scale Height (in kpc) of Runaway Stars Ejected from the Galactic Disk\tablenotemark{a}}
\tablehead{
  \colhead{ \ } &
  \multicolumn{6}{c}{Galactocentric distance $r$}
\\
  \colhead{Stellar Mass} &
  \colhead{$<$ 20 kpc} &
  \colhead{20--40 kpc} &
  \colhead{40--60 kpc} &
  \colhead{60--80 kpc} &
  \colhead{80--100 kpc} &
  \colhead{100--120 kpc}
}
\startdata
1.5 M$_\odot$& 1.5 &  4 & ~9 & 15 & 20 & 25 \\
2   M$_\odot$& 1.5 &  4 & 10 & 15 & 25 & 30 \\
3   M$_\odot$& 1.5 &  4 & 10 & 20 & 30 & 40 \\
4   M$_\odot$& 1.5 &  4 & 15 & 25 & \nodata & \nodata \\
5   M$_\odot$& 2 &  4 & 15 & \nodata & \nodata & \nodata \\
6   M$_\odot$& 2 &  4 & 15 & \nodata & \nodata & \nodata \\
\enddata
\tablenotetext{a}{As in Table \ref{tab:fullmed} for the vertical scale height.}
\label{tab:fullzheight}
\end{deluxetable}

\clearpage

\begin{deluxetable}{ccccccc}
\tablecolumns{8}
\tablewidth{0pc}
\tabletypesize{\footnotesize}
\tablenum{7}
\tablecaption{Median Properties for Unbound Runaway Stars.}
\tablehead{
  \colhead{Stellar Mass} &
  \colhead{$r$ (kpc)} &
  \colhead{\vrad\ (\kms)} &
  \colhead{$b_f$ (deg)}
}
\startdata
1.5 M$_\odot$& 355 & 210 & 9 \\
2   M$_\odot$& 210 & 263 & 9 \\
3   M$_\odot$& ~82 & 360 & 10 \\
4   M$_\odot$& ~43 & 400 & ~9 \\
5   M$_\odot$& ~30 & 410 & ~8 \\
6   M$_\odot$& ~23 & 410 & ~7 \\
\enddata
\label{tab:med-zr}
\end{deluxetable}

\clearpage

\end{document}